\RequirePackage[hyphens]{url}

\documentclass[sigconf,10pt,authorversion]{acmart}

\usepackage{balance}
\usepackage{multirow}
\usepackage{subcaption}
\usepackage{dirtree}
\usepackage[edges]{forest}
\usepackage{booktabs} % For formal tables
\usepackage{colortbl}
\usepackage{graphicx}
\usepackage{enumitem}
\usepackage{arydshln}
\usepackage[final]{listings}
\usepackage[colorinlistoftodos,prependcaption,textsize=small]{todonotes}
\presetkeys%
    {todonotes}%
    {inline}{}

\newcommand{\sys}{Apache Hive}
\newcommand{\mysection}[1]{\vspace{-1.2mm} \section{#1} \vspace{-0.6mm}}
\newcommand{\mysubsection}[1]{\vspace{-1.2mm} \subsection{#1} \vspace{-0.3mm}}

\newcommand{\myparagraph}[1]{\vspace{0.5mm}\noindent\textbf{#1}}
\definecolor{javared}{rgb}{0.6,0,0} % for strings
\definecolor{javagreen}{rgb}{0.25,0.5,0.35} % comments
\definecolor{javapurple}{rgb}{0.5,0,0.35} % keywords
\definecolor{javadocblue}{rgb}{0.25,0.35,0.75} % javadoc

\definecolor{folderbg}{RGB}{124,166,198}
\definecolor{folderborder}{RGB}{110,144,169}
\def\Size{4pt}
\tikzset{%
  folder/.pic={%
    \filldraw [draw=folderborder, top color=folderbg!50, bottom color=folderbg] (-1.05*\Size,0.2\Size+5pt) rectangle ++(.75*\Size,-0.2\Size-5pt);
    \filldraw [draw=folderborder, top color=folderbg!50, bottom color=folderbg] (-1.15*\Size,-\Size) rectangle (1.15*\Size,\Size);
  },
  file/.pic={%
    \filldraw [draw=folderborder, top color=folderbg!5, bottom color=folderbg!10] (-\Size,.4*\Size+5pt) coordinate (a) |- (\Size,-1.2*\Size) coordinate (b) -- ++(0,1.6*\Size) coordinate (c) -- ++(-5pt,5pt) coordinate (d) -- cycle (d) |- (c) ;
  },
}
\forestset{%
  declare autowrapped toks={pic me}{},
  pic dir tree/.style={%
    for tree={%
      folder,
      font=\ttfamily,
      grow'=0,
      edge path={
        \noexpand\path [draw, \forestoption{edge}]
        (!u.south west) +(1pt,0) |- (.child anchor) \forestoption{edge label};
      },
    },
    before typesetting nodes={%
      for tree={%
        edge label+/.option={pic me},
      },
    },
  },
  pic me set/.code n args=2{%
    \forestset{%
      #1/.style={%
        inner sep=1pt,
        inner xsep=7pt,
        pic me={pic {#2}},
      }
    }
  },
  pic me set={directory}{folder},
  pic me set={file}{file},
}

\newcommand*{\equal}{=}

\definecolor{javared}{rgb}{0.6,0,0} % for strings
\definecolor{javagreen}{rgb}{0.25,0.5,0.35} % comments
\definecolor{javapurple}{rgb}{0.5,0,0.35} % keywords
\definecolor{javadocblue}{rgb}{0.25,0.35,0.75} % javadoc

\lstdefinestyle{HIVESQL}{
    language={SQL},
    morekeywords={INT,DECIMAL,TIMESTAMP,VARCHAR,FLOAT,MATERIALIZED,PARTITIONED,STORED,TBLPROPERTIES,POOL,RESOURCE,RULE,WITH,APPLICATION,MAPPING,PLAN,ENABLE,ACTIVATE,TO,MOVE},
    deletekeywords={YEAR},
    deletendkeywords={TIME},
    moredelim=**[is][\btHL]{`}{`},
    moredelim=**[is][{\btHL[fill=green!30,thin]}]{@}{@},
    classoffset=1, % starting new class
    otherkeywords={mat_view},
    morekeywords={mat_view},
    keywordstyle=\bfseries,
    classoffset=0,
}
\lstdefinestyle{JSON}{
    language={JAVA}
}

\copyrightyear{2019} 
\acmYear{2019} 
\setcopyright{acmcopyright}
\acmConference[SIGMOD '19]{2019 International Conference on Management of Data}{June 30-July 5, 2019}{Amsterdam, Netherlands}
\acmBooktitle{2019 International Conference on Management of Data (SIGMOD '19), June 30-July 5, 2019, Amsterdam, Netherlands}
\acmPrice{15.00}
\acmDOI{10.1145/3299869.3314045}
\acmISBN{978-1-4503-5643-5/19/06}

\begin{document}
\title{Apache Hive: From MapReduce to\\Enterprise-grade Big Data Warehousing}

%Add authors as they contribute to article
\author{Jes\'us Camacho-Rodr\'iguez, Ashutosh Chauhan, Alan Gates, Eugene Koifman, Owen O'Malley,}
\author{Vineet Garg, Zoltan Haindrich, Sergey Shelukhin, Prasanth Jayachandran, Siddharth Seth,}
\author{Deepak Jaiswal, Slim Bouguerra, Nishant Bangarwa, Sankar Hariappan, Anishek Agarwal,}
\author{Jason Dere, Daniel Dai, Thejas Nair, Nita Dembla, Gopal Vijayaraghavan, G\"{u}nther Hagleitner}
\affiliation{\vskip0.35ex Hortonworks Inc.\\[0.5ex]}

\begin{abstract}
\sys{} is an open-source relational database system for analytic big-data workloads. In this paper we describe the key innovations on the journey from batch tool to fully fledged enterprise data warehousing system. We present a hybrid architecture that combines traditional MPP techniques with more recent big data and cloud concepts to achieve the scale and performance required by today's analytic applications. We explore the system by detailing enhancements along four main axis: Transactions, optimizer, runtime, and federation. We then provide experimental results to demonstrate the performance of the system for typical workloads and conclude with a look at the community roadmap.
\end{abstract}


%\renewcommand{\shorttitle}{}
%\renewcommand{\shortauthors}{}

\maketitle

\mysection{Introduction}
%All authors

When Hive was first introduced over 10 years ago~\cite{DBLP:conf/icde/ThusooSJSCZALM10}, the motivation of the authors was to expose a SQL-like interface on top of Hadoop MapReduce to abstract the users from dealing with low level implementation details for their parallel batch processing jobs. Hive focused mainly on Extract-Transform-Load (ETL) or batch reporting workloads that consisted of ($i$)~reading huge amounts of data, ($ii$)~executing transformations over that data (e.g., data wrangling, consolidation, aggregation) and finally ($iii$)~loading the output into other systems that were used for further analysis.

As Hadoop became a ubiquitous platform for inexpensive data storage with HDFS, developers focused on increasing the range of workloads that could be executed efficiently within the platform. YARN~\cite{DBLP:conf/cloud/VavilapalliMDAKEGLSSSCORRB13}, a resource management framework for Hadoop, was introduced, and shortly afterwards, data processing engines (other than MapReduce) such as Spark~\cite{DBLP:conf/hotcloud/ZahariaCFSS10,website:Spark} or Flink~\cite{DBLP:journals/debu/CarboneKEMHT15,website:Flink} were enabled to run on Hadoop directly by supporting YARN.

Users also increasingly focused on migrating their data warehousing workloads from other systems to Hadoop. These workloads included interactive and ad-hoc reporting, dashboarding and other business intelligence use cases. There was a common requirement that made these workloads a challenge on Hadoop: They required a \textit{low-latency SQL engine}. Multiple efforts to achieve this were started in parallel and new SQL MPP systems compatible with YARN such as Impala~\cite{DBLP:conf/cidr/KornackerBBBCCE15,website:Impala} and Presto~\cite{website:Presto} emerged.

Instead of implementing a new system, the Hive community concluded that the current implementation of the project provided a good foundation to support these workloads. Hive had been designed for large-scale reliable computation in Hadoop, and it already provided SQL compatibility (alas, limited) and connectivity to other data management systems. However, Hive needed to evolve and undergo major renovation to satisfy the requirements of these new use cases, adopting common data warehousing techniques that had been extensively studied over the years.

Previous works presented to the research community about Hive focused on ($i$)~its initial architecture and implementation on top of HDFS and MapReduce~\cite{DBLP:conf/icde/ThusooSJSCZALM10}, and ($ii$)~improvements to address multiple performance shortcomings in the original system, including the introduction of an optimized columnar file format, physical optimizations to reduce the number of MapReduce phases in query plans, and a vectorized execution model to improve runtime efficiency~\cite{DBLP:conf/sigmod/HuaiCGHHOPYL014}. Instead, this paper describes the significant novelties introduced in Hive after the last article was presented. In particular, it focuses on the renovation work done to improve the system along four different axes:

\vspace{1.1mm}
\myparagraph{SQL and ACID support (Section~\ref{sec:sqlacid}).} SQL compliance is a key requirement for data warehouses. Thus, SQL support in Hive has been extended, including correlated subqueries, integrity constraints, and extended OLAP operations, among others. In turn, warehouse users need support to \textit{insert}, \textit{update}, \textit{delete}, and \textit{merge} their individual records on demand. Hive provides ACID guarantees with Snapshot Isolation using a transaction manager built on top of the Hive Metastore.

\myparagraph{Optimization techniques (Section~\ref{sec:optimization}).} Query optimization is especially relevant for data management systems that use declarative querying languages such as SQL. Instead of implementing its own optimizer from scratch, Hive chooses to integrate with Calcite~\cite{DBLP:conf/sigmod/BegoliCHML18,website:Calcite} and bring its optimization capabilities to the system. In addition, Hive includes other optimization techniques commonly used in data warehousing environments such as query reoptimization, query result caching, and materialized view rewriting.

\myparagraph{Runtime latency (Section~\ref{sec:execution}).} To cover a wider variety of use cases including interactive queries, it is critical to improve latency. Hive support for optimized columnar data storage and vectorization of operators were presented in a previous work~\cite{DBLP:conf/sigmod/HuaiCGHHOPYL014}. In addition to those improvements, Hive has since moved from MapReduce to Tez~\cite{DBLP:conf/sigmod/SahaSSVMC15,website:Tez}, a YARN compatible runtime that provides more flexibility to implement arbitrary data processing applications than MapReduce. Furthermore, Hive includes LLAP, an additional layer of persistent long-running executors that provides data caching, facilities runtime optimizations, and avoids YARN containers allocation overhead at start-up.

\myparagraph{Federation capabilities (Section~\ref{sec:fed}).} One of the most important features of Hive is the capacity to provide a unified SQL layer on top of many specialized data management systems that have emerged over the last few years. Thanks to its Calcite integration and storage handler improvements, Hive can seamlessly push computation and read data from these systems. In turn, the implementation is easily extensible to support other systems in the future.
\vspace{1.1mm}

The rest of the paper is organized as follows. Section~\ref{sec:arch} provides background on Hive's architecture and main components. Sections~\ref{sec:sqlacid}-\ref{sec:fed} describe our main contributions to improve the system along the axes discussed above. Section~\ref{sec:exp} presents an experimental evaluation of Hive. Section~\ref{sec:discussion} discusses the impact of the new features, while Section~\ref{sec:roadmap} briefly describes the roadmap for the project. Finally, Section~\ref{sec:conclusion} summarizes our conclusions.

%%%%%SECTIONS%%%%%
%%%%%SECTION%%%%%
\mysection{System architecture}
\label{sec:arch}

\begin{figure}[t]
\centering
\includegraphics[width=\columnwidth]{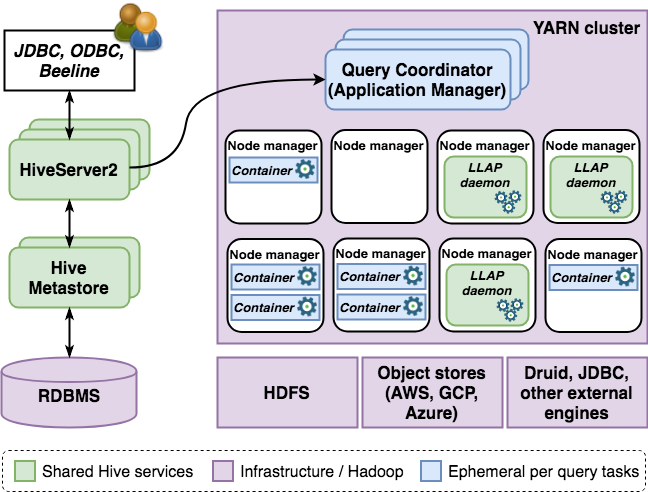}
\caption{Apache Hive architecture.\label{fig:arch}}
\end{figure}

\begin{figure*}[t]
\centering
\includegraphics[width=0.94\textwidth]{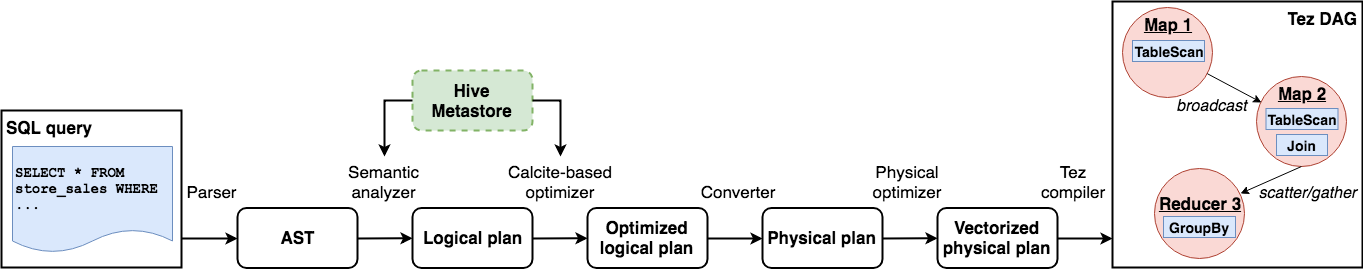}
\vspace{-1mm}\caption{Query preparation stages in HiveServer2.\label{fig:stages}}
\end{figure*}

In this section we briefly introduce Hive's architecture. Figure~\ref{fig:arch} depicts the main components in the system.

\myparagraph{Data storage.} Data in Hive can be stored using any of the supported file formats in any file system compatible with Hadoop. As of today, the most common file formats are ORC~\cite{website:Orc} and Parquet~\cite{website:Parquet}. In turn, compatible file systems include HDFS which is the most commonly used distributed file system implementation, and all the major commercial cloud object stores such as AWS S3 and Azure Blob Storage. In addition, Hive can also read and write data to other standalone processing systems, such as Druid~\cite{DBLP:conf/sigmod/YangTLRMG14,website:Druid} or HBase~\cite{website:HBase}, which we discuss in more detail in Section~\ref{sec:fed}.

\myparagraph{Data catalog.} Hive stores all information about its data sources using the Hive Metastore (or HMS, in short). In a nutshell, HMS is a catalog for all data queryable by Hive. It uses a RDBMS to persist the information, and it relies on DataNucleus~\cite{website:DataNucleus}, a Java object-relational mapping implementation, to simplify the support of multiple RDBMS at the backend. For calls that require low latency, HMS may bypass DataNucleus and query the RDMBS directly. The HMS API supports multiple programming languages and the service is implemented using Thrift~\cite{website:Thrift}, a software framework that provides an interface definition language, code generation engine, and binary communication protocol implementation.

\myparagraph{Exchangeable data processing runtime.} Hive has become one of the most popular SQL engines on top of Hadoop and it has gradually moved away from MapReduce to support more flexible processing runtimes compatible with YARN~\cite{DBLP:conf/sigmod/SahaSSVMC15}. While MapReduce is still supported, currently the most popular runtime for Hive is Tez~\cite{DBLP:conf/sigmod/SahaSSVMC15,website:Tez}. Tez provides more flexibility than MapReduce by modeling data processing as DAGs with vertices representing application logic and edges representing data transfer, similar to other systems such as Dryad~\cite{DBLP:conf/eurosys/IsardBYBF07} or Hyracks~\cite{DBLP:conf/icde/BorkarCGOV11}. In addition, Tez is compatible with LLAP, the persistent execution and cache layer presented in Section~\ref{sec:execution}.

\myparagraph{Query server.} HiveServer2 (or HS2, in short) allows users to execute SQL queries in Hive. HS2 supports local and remote JDBC and ODBC connections; Hive distribution includes a JDBC thin client called Beeline.%~\footnote{HiveServer2 and Beeline replace the original Hive Command Line Interface (CLI) thick client built for Hive on Hadoop, which included HiveServer1.}.

Figure~\ref{fig:stages} depicts the stages that a SQL query goes through in HS2 to become an executable plan. Once a user submits a query to HS2, the query is handled by the driver, which parses the statement and generates a Calcite~\cite{DBLP:conf/sigmod/BegoliCHML18,website:Calcite} logical plan from its AST.
The Calcite plan is then optimized. Note that HS2 accesses information about the data sources in HMS for validation and optimization purposes. Subsequently, the plan is converted into a physical plan, potentially introducing additional operators for data partitioning, sorting, etc. HS2 executes additional optimizations on the physical plan DAG, and if all operators and expressions in the plan are supported, a vectorized plan~\cite{DBLP:conf/sigmod/HuaiCGHHOPYL014} may be generated from it.
The physical plan is passed to the task compiler, which breaks the operator tree into a DAG of executable tasks. Hive implements an individual task compiler for each supported processing runtime, i.e., Tez, Spark, and MapReduce.
After the tasks are generated, the driver submits them to the runtime application manager in YARN, which handles the execution. For each task, the physical operators within that task are first initialized and then they process the input data in a pipelined fashion. After execution finishes, the driver fetches the results for the query and returns them to the user.

%Note that during SQL compilation, HS2 accesses information about the data sources in HMS for validation and optimization purposes.

%%%%%SECTION%%%%%
\mysection{SQL and ACID support}
\label{sec:sqlacid}
Standard SQL and ACID transactions are critical requirements in enterprise data warehouses. In this section, we present Hive's broaden SQL support. Additionally, we describe the improvements made to Hive in order to provide ACID guarantees on top of Hadoop.

\mysubsection{SQL support}
\label{subsec:sql}

\begin{figure}[t]
\scriptsize
\begin{forest}
  pic dir tree,
  where level=0{}{% folder icons by default; override using file for file icons
    directory,
  },
[\hspace{-10pt}\underline{warehouse}
  [database
    [store\_sales
      [sold\_date\_sk\equal{}1
        [base\_100
          [file\_0000, file]
          [file\_0001, file]
        ]
        [delta\_101\_105
          [file\_0000, file]
          [file\_0001, file]
        ]
        [delete\_delta\_103\_103
          [file\_0000, file]
          [file\_0001, file]
        ]
      ]
      [sold\_date\_sk\equal{}2
        [\ldots\phantom{others}]
      ]
      [\ldots\phantom{others}]
    ]
  ]
]
\end{forest}
\vspace{-1mm}\caption{Physical layout for partitioned table.\label{fig:partitionedtable}}\vspace{-2mm}
\end{figure}
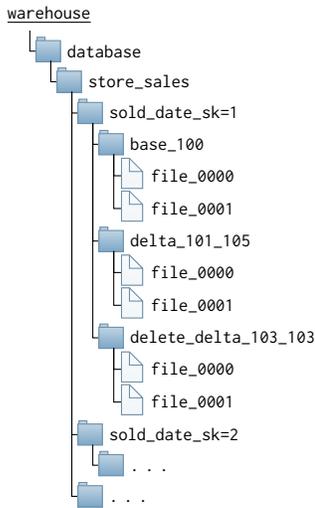

To provide a replacement for traditional data warehouses, Hive needed to be extended to support more features from standard SQL.
Hive uses a nested data model supporting all major atomic SQL data types as well as non-atomic types such as \texttt{STRUCT}, \texttt{ARRAY} and \texttt{MAP}.
Besides, each new Hive release has increased its support for important constructs that are part of the SQL specification. For instance, there is extended support for correlated subqueries, i.e., subqueries that reference columns from the outer query, advanced OLAP operations such as grouping sets or window functions, set operations, and integrity constraints, among others.
On the other hand, Hive has preserved multiple features of its original query language that were valuable for its user base. One of the most popular features is being able to specify the physical storage layout at table creation time using a \texttt{PARTITIONED BY} columns clause. In a nutshell, the clause lets a user partition a table horizontally. Then Hive stores the data for each set of partition values in a different directory in the file system. To illustrate the idea, consider the following table definition and the corresponding physical layout depicted in Figure~\ref{fig:partitionedtable}:

\begin{lstlisting}[style=HIVESQL]
CREATE TABLE store_sales (
  sold_date_sk INT, item_sk INT, customer_sk INT, store_sk INT,
  quantity INT, list_price DECIMAL(7,2), sales_price DECIMAL(7,2)
) PARTITIONED BY (sold_date_sk INT);
\end{lstlisting}

The advantage of using the \texttt{PARTITIONED BY} clause is that Hive will be able to skip scanning full partitions easily for queries that filter on those values.

%From 2010 original Hive paper:
%However, we do realize that with more frequent loads the number of partitions can become very large and that may require us to implement INSERT INTO semantics. The lack of INSERT INTO, UPDATE and DELETE in Hive on the other hand do allow us to use very simple mechanisms to deal with reader and writer concurrency without implementing complex locking protocols.

\mysubsection{ACID implementation}
\label{subsec:acid}

Initially, Hive only had support to insert and drop full partitions from a table~\cite{DBLP:conf/icde/ThusooSJSCZALM10}. Although the lack of row level operations was acceptable for ETL workloads, as Hive evolved to support many traditional data warehousing workloads, there was an increasing requirement for full DML support and ACID transactions. Hence, Hive now includes support to execute \texttt{INSERT}, \texttt{UPDATE}, \texttt{DELETE}, and \texttt{MERGE} statements. It provides ACID guarantees via Snapshot Isolation~\cite{DBLP:conf/sigmod/CahillRF08} on read and well defined semantics in case of failure using a transaction manager built on top of the HMS. Currently transactions can only span a single statement; we plan to support multi-statement transactions in the near future. However, it is possible to write to multiple tables within a single transaction using Hive multi-insert statements~\cite{DBLP:conf/icde/ThusooSJSCZALM10}.

The main challenges to overcome in supporting row level operations in Hive were ($i$)~the lack of a transaction manager in the system, and ($ii$)~the lack of file updates support in the underlying file system. In the following, we provide more details about the implementation of ACID in Hive and how these issues were addressed.

\myparagraph{Transaction and lock management.} Hive stores transaction and locking information state in HMS. It uses a global transaction identifier or \textit{TxnId}, i.e., a monotonically increasing value generated by the Metastore, for each transaction run in the system. In turn, each \textit{TxnId} maps into one or multiple write identifiers or \textit{WriteIds}. A \textit{WriteId} is a monotonically increasing value generated by the Metastore as well, but within a table scope. The \textit{WriteId} is stored with each record that is written by a transaction; all records written by the same transaction to the same table share the same \textit{WriteId}. In turn, files sharing the same \textit{WriteId} are identified uniquely using a \textit{FileId}, while each record within a file is identified uniquely by a \textit{RowId} field. Note that the combination of \textit{WriteId}, \textit{FileId} and \textit{RowId} identifies uniquely each record in a table. A delete operation in Hive is modeled as an insert of a labeled record that points to the unique identifier of the record being deleted.

To achieve Snapshot Isolation, HS2 obtains a \textit{logical snapshot} of the data that needs to read when a query is executed. The snapshot is represented by a transaction list comprising the highest allocated \textit{TxnId} at that moment, i.e., the high watermark, and the set of open and aborted transactions below it. For each table that the query needs to read, HS2 first generates the \textit{WriteId} list from the transaction list by contacting HMS; the \textit{WriteId} list is similar to the transaction list but within the scope of a single table. Each scan operation in the plan is bound to a \textit{WriteId} list during compilation. The readers in that scan will skip rows whose \textit{WriteId} ($i$)~is higher than the high watermark, or ($ii$)~is part of the set of open and aborted transactions. The reason to keep both global and per-table identifiers is that the readers for each table keep a smaller state, which becomes critical for performance when there are a large number of open transactions in the system.

For partitioned tables the lock granularity is a partition, while the full table needs to locked for unpartitioned tables.
HS2 only needs to obtain exclusive locks for operations that disrupt readers and writers, such as a \texttt{DROP PARTITION} or \texttt{DROP TABLE} statements. All other common operations just acquire shared locks. Updates and deletes use optimistic conflict resolution by tracking their write sets and resolving the conflict at commit time, letting the first commit win.

\myparagraph{Data and file layout.} Hive stores data for each table and partition in a different directory (recall Figure~\ref{fig:partitionedtable}). Similar to~\cite{DBLP:conf/sigmod/LarsonCFHMNPPRRS13}, we use different stores or directories within each table or partition to support concurrent read and write operations: \textit{base} and \textit{delta}, which in turn may contain one or multiple files. The files in the base store contain all valid records up to a certain \textit{WriteId}. For instance, folder \texttt{base\_100} contains all records up to \textit{WriteId} $100$. On the other hand, a delta directory contains files with records within a \textit{WriteId} range. Hive keeps separate delta directories for inserted and deleted records; update operations are split into delete and insert operations. An insert or delete transaction creates a delta directory with records bound to a single \textit{WriteId}, e.g., \texttt{delta\_101\_101} or \texttt{delete\_delta\_102\_102}. Delta directories containing more than one \textit{WriteId} are created as part of the compaction process (explained below).

As mentioned previously, a table scan in a query has a \textit{WriteId} list associated to it. The readers in the scan discard full directories as well as individual records that are not valid based on the current snapshot. When deletes are present in the delta files, records in the base and insert delta files need to be anti-joined with the delete deltas that apply to their \textit{WriteId} range. Since delta files with deleted records are usually small, they can be kept in-memory most times, accelerating the merging phase.

\myparagraph{Compaction.} Compaction is the process in Hive that merges files in delta directories with other files in delta directories (referred to as \textit{minor compaction}), or files in delta directories with files in base directories (referred to as \textit{major compaction}). The crucial reasons to run compaction periodically are ($i$)~decreasing the number of directories and files in tables, which could otherwise affect file system performance, ($ii$)~reducing the readers effort to merge files at query execution time, and ($iii$)~shorten the set of open and aborted {TxnIds} and \textit{WriteIds} associated with each snapshot, i.e., major compaction deletes history, increasing the \textit{TxnId} under which all records in the tables are known to be valid.

Compaction is triggered automatically by HS2 when certain thresholds are surpassed, e.g., number of delta files in a table or ratio of records in delta files to base files. Finally, note that compaction does not need any locks over the table. In fact, the cleaning phase is separated for the merging phase so that any ongoing query can complete its execution before files are deleted from the system.

%%%%%SECTION%%%%%
\mysection{Query optimization}
\label{sec:optimization}

While the support for optimization in the initial versions of Hive was limited, it is evident that the development of its execution internals is not sufficient to guarantee efficient performance. Thus, currently the project includes many of the complex techniques typically used in relational database systems. This section describes the most important optimization features that help the system generate better plans and deliver improvements to query execution, including its fundamental integration with Apache Calcite.

\mysubsection{Rule and cost-based optimizer}
\label{subsec:rbcbopt}

Initially, Apache Hive executed multiple rewritings to improve performance while parsing the input SQL statement. In addition, it contained a rule-based optimizer that applied simple transformations to the physical plan generated from the query. For instance, the goal of many of these optimizations was trying to minimize the cost of data shuffling, a critical operation in the MapReduce engine. There were other optimizations to push down filter predicates, project unused columns, and prune partitions. While this was effective for some queries, working with the physical plan representation made implementing complex rewritings such as join reordering, predicate simplification and propagation, or materialized view-based rewriting, overly complex.

For that reason, a new plan representation and optimizer powered by Apache Calcite~\cite{website:Calcite,DBLP:conf/sigmod/BegoliCHML18} were introduced. Calcite is a modular and extensible query optimizer with built-in elements that can be combined in different ways to build your own optimization logic. These include diverse rewriting rules, planners, and cost models.

Calcite provides two different planner engines: ($i$)~a \textit{cost-based planner}, which triggers rewriting rules with the goal of reducing the overall expression cost, and ($ii$)~an \textit{exhaustive planner}, which triggers rules exhaustively until it generates an expression that is no longer modified by any rules. Transformation rules work indistinctly with both planners.

Hive implements \textit{multi-stage optimization} similar to other query optimizers~\cite{DBLP:conf/sigmod/SolimanAREGSCGRPWNKB14}, where each optimization stage uses a planner and a set of rewriting rules. This allows Hive to reduce the overall optimization time by guiding the search for different query plans. Some of the Calcite rules enabled in Apache Hive are join reordering, multiple operators reordering and elimination, constant folding and propagation, and constraint-based transformations.

\myparagraph{Statistics.} Table statistics are stored in the HMS and provided to Calcite at planning time. These include the table cardinality, and number of distinct values, minimum and maximum value for each column. The statistics are stored such that they can be combined in an additive fashion, i.e., future inserts as well as data across multiple partitions can add onto existing statistics. The range and cardinality are trivially mergeable. For the number of distinct values, HMS uses a bit array representation based on HyperLogLog++~\cite{DBLP:conf/edbt/HeuleNH13} which can be combined without loss of approximation accuracy.

\mysubsection{Query reoptimization}
\label{subsec:reopt}

Hive supports query reoptimization when certain errors are thrown during execution. In particular, it implements two independent reoptimization strategies.

The first strategy, \textit{overlay}, changes certain configuration parameters for all query reexecutions. For instance, a user may choose to force all joins in query reexecutions to use a certain algorithm, e.g., \textit{hash partitioning with sort-merge}. This may be useful when certain configuration values are known to make query execution more robust.

The second strategy, \textit{reoptimize}, relies on statistics captured at runtime. While a query is being planned, the optimizer estimates the size of the intermediate results in the plan based on statistics that are retrieved from HMS. If those estimates are not accurate, the optimizer may make planning mistakes, e.g., wrong join algorithm selection or memory allocation. This in turn may lead to poor performance and execution errors. Hive captures runtime statistics for each operator in the plan. If any of the aforementioned problems is detected during query execution, the query is reoptimized using the runtime statistics and executed again.

\mysubsection{Query results cache}
\label{subsec:querycache}

Transactional consistency of the warehouse allows Hive to reuse the results of a previously executed query by using the internal transactional state of the participating tables. The query cache provides scalability advantages when dealing with BI tools which generate repetitive identical queries.

Each HS2 instance keeps its own \textit{query cache} component, which in turn keeps a map from the query AST representation to an entry containing the results location and information of the snapshot of the data over which the query was answered. This component is also responsible for expunging stale entries and cleaning up resources used by those entries.

During query compilation, HS2 checks its cache using the input query AST in a preliminary step. The unqualified table references in the query are resolved before the AST is used to prove the cache, since depending on the current database at query execution time, two queries with the same text may access tables from different databases. If there is a cache hit and the tables used by the query do not contain new or modified data, then the query plan will consist on a single task that will fetch the results from the cached location. If the entry does not exist, then the query is run as usual and results generated for the query are saved to the cache if the query meets some conditions; for instance, the query cannot contain non-deterministic functions (\textit{rand}), runtime constant functions (\textit{current\_date}, \textit{current\_timestamp}), etc.

The query cache has a pending entry mode, which protects against a thundering herd of identical queries when data is updated and a cache miss is observed by several of them at the same time. The cache will be refilled by the first query that comes in. Besides, that query may potentially receive more cluster capacity since it will serve results to every other concurrent identical query suffering a cache miss.

\mysubsection{Materialized views and rewriting}
\label{subsec:matviews}

Traditionally, one of the most powerful techniques used to accelerate query processing in data warehouses is the pre-computation of relevant materialized views~\cite{DBLP:conf/icde/GuptaHRU97,DBLP:conf/icdt/Gupta97,DBLP:conf/sigmod/HarinarayanRU96,DBLP:conf/icde/ChaudhuriKPS95, DBLP:conf/sigmod/GoldsteinL01}.

%\centering{
%\begin{tikzpicture}
%[level/.style={level distance=10mm}]
%\tikzstyle{edge from parent}=[opacity=1,thick,draw]
%\tikzstyle{level 1}=[sibling distance=2.1cm,level distance=-7mm]
%\tikzstyle{level 2}=[sibling distance=2.2cm]
%\node[pname] {$q_1$}
%	child {
%          node{~} edge from parent [transparent edge]
%	}
%		child {
%          node{~} edge from parent [transparent edge]
%             child{node[plain] {site} edge from parent [descendant]	
%               child{node[plain] {people} edge from parent [child]	
%                 child{node[plain] {person} edge from parent [child]
%                   child{node[plain] {@id$_{=person0}$} edge from parent [child]}
%	               child {node[plain] {name$_{val}$} edge from parent [child]}
%	             }
%	           }
%           }
%	}
%;
%\end{tikzpicture}}
\begin{figure}[t]
\centering{
\setlength{\tabcolsep}{0.75ex}%
\begin{tabular}{|l|l|}
\hline
\small{\textbf{(a)}}
& {\begin{lstlisting}[style=HIVESQL,numbers=none,xleftmargin=0pt]
CREATE MATERIALIZED VIEW mat_view AS
SELECT d_year, d_moy, d_dom, SUM(ss_sales_price) AS sum_sales
FROM store_sales, date_dim
WHERE ss_sold_date_sk = d_date_sk AND d_year > 2017
GROUP BY d_year, d_moy, d_dom;
\end{lstlisting}} \\[4.25ex]
\hline\hline
\multirow{7}{*}{\small{\textbf{(b)}}} &
\footnotesize{$q_1$:} \\
& {\hspace{1ex}\begin{lstlisting}[style=HIVESQL,numbers=none,xleftmargin=0pt]
SELECT SUM(ss_sales_price) AS sum_sales
FROM store_sales, date_dim
WHERE ss_sold_date_sk = d_date_sk AND
  d_year = 2018 AND d_moy IN (1,2,3);
\end{lstlisting}} \\[2ex]
& \footnotesize{$q'_1$:} \\
& {\hspace{1ex}\begin{lstlisting}[style=HIVESQL,numbers=none,xleftmargin=0pt]
SELECT SUM(sum_sales)
FROM mat_view
WHERE d_year = 2018 AND d_moy IN (1,2,3);
\end{lstlisting}} \\[2.25ex]
\hline\hline
\multirow{14}{*}{\small{\textbf{(c)}}} &
\footnotesize{$q_2$:} \\
& {\hspace{1ex}\begin{lstlisting}[style=HIVESQL,numbers=none,xleftmargin=0pt]
SELECT d_year, d_moy, SUM(ss_sales_price) AS sum_sales
FROM store_sales, date_dim
WHERE ss_sold_date_sk = d_date_sk AND d_year > 2016
GROUP BY d_year, d_moy;
\end{lstlisting}} \\[2ex]
& \footnotesize{$q'_2$:} \\
& {\hspace{1ex}\begin{lstlisting}[style=HIVESQL,numbers=none,xleftmargin=0pt]
SELECT d_year, d_moy, SUM(sum_sales)
FROM (
  SELECT d_year, d_moy, SUM(sum_sales) AS sum_sales
  FROM mat_view
  GROUP BY d_year, d_moy
  UNION ALL
  SELECT d_year, d_moy, SUM(ss_sales_price) AS sum_sales
  FROM store_sales, date_dim
  WHERE ss_sold_date_sk = d_date_sk AND
    d_year > 2016 AND d_year <= 2017
  GROUP BY d_year, d_moy
) subq
GROUP BY d_year, d_moy;
\end{lstlisting}} \\[12.75ex]
\hline
\end{tabular}}
\vspace{-2mm}\caption{Materialized view definition (a) and sample fully (b) and partially (c) contained rewritings.\label{fig:mv_rewriting}}\vspace{-2mm}
\end{figure}

Apache Hive supports materialized views and automatic query rewriting based on those materializations. In particular, materialized views are just \textit{semantically enriched} tables. Therefore, they can be stored natively by Hive or in other supported systems (see Section~\ref{sec:fed}), and they can seamlessly exploit features such as LLAP acceleration (described in Section~\ref{subsec:llap}). The optimizer relies in Calcite to automatically produce full and partially contained rewritings on Select-Project-Join-Aggregate (SPJA) query expressions (see Figure~\ref{fig:mv_rewriting}). The rewriting algorithm exploits integrity constraints information declared in Hive, e.g., \textit{primary key}, \textit{foreign key}, \textit{unique key}, and \textit{not null}, to produce additional valid transformations. The algorithm is encapsulated within a rule and it is triggered by the cost-based optimizer, which is responsible to decide whether a rewriting should be used to answer a query or not. Note that if multiple rewritings are applicable to different parts of a query, the optimizer may end up selecting more than one view substitutions.

\myparagraph{Materialized view maintenance.} When data in the sources tables used by a materialized view changes, e.g., new data is inserted or existing data is modified, we will need to refresh the contents of the materialized view to keep it up-to-date with those changes. Currently, the rebuild operation for a materialized view needs to be triggered manually by the user using a \texttt{REBUILD} statement.

By default, Hive attempts to rebuild a materialized view incrementally~\cite{DBLP:journals/debu/GuptaM95,DBLP:conf/sigmod/GriffinL95}, falling back to full rebuild if it is not possible. Current implementation only supports incremental rebuild when there were \texttt{INSERT} operations over the source tables, while \texttt{UPDATE} and \texttt{DELETE} operations will force a full rebuild of the materialized view.

One interesting aspect is that the incremental maintenance relies on the rewriting algorithm itself. Since a query is associated with a snapshot of the data, the materialized view definition is enriched with filter conditions on the \textit{WriteId} column value of each table scanned (recall Section~\ref{subsec:acid}). Those filters conditions reflect the snapshot of the data when the materialized view was created or lastly refreshed. When the maintenance operation is triggered, the rewriting algorithm may produce a partially contained rewriting that reads the materialized view and the new data from the source tables. This rewritten plan is in turn transformed into ($i$)~an \texttt{INSERT} operation if it is a SPJ materialized view, or ($ii$)~a \texttt{MERGE} operation if it is a SPJA materialized view.

\myparagraph{Materialized view lifecycle.} By default, once the materialized view contents are stale, the materialized view will not be used for query rewriting.

However, in some occasions it may be fine to accept rewriting on stale data while updating the materialized views in micro batches. For those cases, Hive lets users combine a rebuild operation run periodically, e.g., every 5~minutes, and define a window for data staleness allowed in the materialized view definition using a table property~\footnote{Table properties allow users in Hive to tag table and materialized view definitions with metadata key-value pairs.}, e.g., 10~minutes.

\mysubsection{Shared work optimization}
\label{subsec:swo}

Hive is capable of identifying overlapping subexpressions within the execution plan of a given query, computing them only once and reusing their results. Instead of triggering transformations to find equivalent subexpressions within the plan, the \textit{shared work optimizer} only merges \textit{equal} parts of a plan, similar to other reuse-based approaches presented in prior works~\cite{DBLP:conf/cikm/Camacho-Rodriguez16,DBLP:conf/sigmod/JindalQPYDBFLKR18}. It applies the reutilization algorithm just before execution: it starts merging scan operations over the same tables, then it continues merging plan operators until a difference is found. The decision on the data transfer strategy for the new shared edge coming out of the merged expression is left to the underlying engine, i.e., Apache Tez. The advantage of a reuse-based approach is that it can accelerate execution for queries computing the same subexpression more than once, without introducing much optimization overhead. Nonetheless, since the shared work optimizer does not explore the complete search space of equivalent plans, Hive may fail to detect existing reutilization opportunities.

\mysubsection{Dynamic semijoin reduction}
\label{subsec:semijoin}

Semijoin reduction is a traditional technique used to reduce the size of intermediate results during query execution~\cite{DBLP:journals/tods/BernsteinGWRR81,DBLP:journals/tse/ApersHY83}. The optimization is specially useful for star schema databases with one or many dimension tables. Queries on those databases usually join the fact table and dimension tables, which are filtered with predicates on one or multiple columns. However, those columns are not used in the join condition, and thus, a filter over the fact table cannot be created statically. The following SQL query shows an example of such a join between the \textit{store\_sales} and \textit{store\_returns} fact tables and the \textit{item} dimension table:

\begin{lstlisting}[style=HIVESQL,label=lst:semijoin]
SELECT ss_customer_sk, SUM(ss_sales_price) AS sum_sales
FROM store_sales, store_returns, item
WHERE ss_item_sk = sr_item_sk AND
  ss_ticket_number = sr_ticket_number AND
  ss_item_sk = i_item_sk AND
  i_category = 'Sports'
GROUP BY ss_customer_sk
ORDER BY sum_sales DESC;
\end{lstlisting}

By applying semijoin reduction, Hive evaluates the subexpression that is filtered (in the example above, the filter on the \textit{item} table), and subsequently, the values produced by the expression are used to skip reading records from the rest of tables.

The semijoin reducers are introduced by the optimizer and they are pushed into the scan operators in the plan. Depending on the data layout, Hive implements two variants of the optimization.

\myparagraph{Dynamic partition pruning.} It is applied if the table reduced by the semijoin is partitioned by the join column. Once the filtering subexpression is evaluated, Hive uses the values produced to skip reading unneeded partitions dynamically, while the query is running. Recall that each table partition value is stored in a different folder in the file system, hence skipping them is straightforward.

\myparagraph{Index semijoin.} If the table reduced by the semijoin is not partitioned by the join column, Hive may use the values generated by the filtering subexpression to ($i$)~create a range filter condition with the minimum and maximum values, and ($ii$)~create a Bloom filter with the values produced by the subexpression. Hive populates the semijoin reducer with these two filters, which may be used to avoid scanning entire row groups at runtime, e.g., if data is stored in ORC files~\cite{DBLP:conf/sigmod/HuaiCGHHOPYL014}.

%Jesus: I am going to comment this out for the time being, since we are short in space (~12 pages) and I assume reader will be familiar with SJ optimization. If we have additional space, we may add it back later.
%For the query mentioned earlier~\ref{lst:semijoin}, the semijoin reduction results in a plan which is not a tree, where the reduction forms a dependency acyclic graph which is satisfied at runtime.
%\includegraphics[width=\columnwidth]{figures/bloom2.png}

%%%%%SECTION%%%%%
\mysection{Query execution}
\label{sec:execution}

Improvements in the query execution internals such as the transition from MapReduce to Apache Tez~\cite{DBLP:conf/sigmod/SahaSSVMC15} and the implementation of columnar-based storage formats and vectorized operators~\cite{DBLP:conf/sigmod/HuaiCGHHOPYL014} reduced query latency in Hive by orders of magnitude. However, to improve execution runtime further, Hive needed additional enhancements to overcome limitations inherent to its initial architecture that was tailored towards long running queries. Among others, ($i$)~execution required YARN containers allocation at start-up, which quickly became a critical bottleneck for low latency queries, ($ii$)~Just-In-Time (JIT) compiler optimizations were not as effective as possible because containers were simply killed after query execution, and ($iii$)~Hive could not exploit data sharing and caching possibilities within and among queries, leading to unnecessary IO overhead.

\mysubsection{LLAP: Live Long and Process}
\label{subsec:llap}

\textit{Live Long and Process}, also known as LLAP, is an optional layer that provides persistent multi-threaded query executors and multi-tenant in-memory cache to deliver faster SQL processing at large scale in Hive. LLAP does not replace the existing execution runtime used by Hive, such as Tez, but rather enhances it. In particular, execution is scheduled and monitored by Hive query coordinators transparently over both LLAP nodes as well as regular containers. 

The data I/O, caching, and query fragment execution capabilities of LLAP are encapsulated within \textit{daemons}. Daemons are setup to run continuously in the worker nodes in the cluster, facilitating JIT optimization, while avoiding any start-up overhead. YARN is used for coarse-grained resource allocation and scheduling. It reserves memory and CPU for the daemons and handles restarts and relocation. The daemons are stateless: each contains a number of executors to run several query fragments in parallel and a local work queue. Failure and recovery is simplified because any node can still be used to process any fragment of the input data if a LLAP daemon fails.

\myparagraph{I/O elevator.} The daemon uses separate threads to off-load data I/O and decompression, which we refer to as \textit{I/O elevator}. Data is read in batches and transformed into an internal run-length encoded (RLE) columnar format ready for vectorized processing. A column batch is moved into execution phase as soon as it is read, which allows previous batches to be processed while the following batches are being prepared.

Transformation from underlying file format into LLAP internal data format is accomplished using plugins that are specific to each format. Currently LLAP supports translation from ORC~\cite{website:Orc}, Parquet~\cite{website:Parquet}, and text file formats.

The I/O elevator can push down projections, sargable predicates~\footnote{Predicates that can be evaluated using an index seek.} and Bloom filters to the file reader, if they are provided. For instance, ORC files can take advantage of these structures to skip reading entire column and row groups.

\begin{figure}[t]
\centering
\includegraphics[width=0.7\columnwidth]{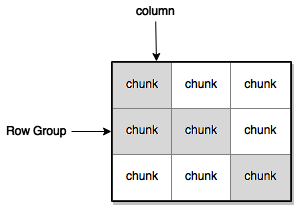}
\vspace{-2mm}\caption{Cache addressing in LLAP.}\label{fig:cache_chunks}\vspace{-2mm}
\end{figure}

\myparagraph{Data caching.} LLAP features an off-heap cache as its primary buffer pool for holding data coming into the I/O elevator. For each file, the cache is addressed by the I/O elevator along two dimensions: row and column groups. A set of rows and columns form a row-column \textit{chunk} (refer to Figure~\ref{fig:cache_chunks}). The I/O elevator reassembles the selected projection and evaluates the predicates to extract chunks which can be reconstituted into vector batches for the operator pipeline. In case of cache misses, the cache is repopulated with the missing chunks before the reconstitution is performed. The result is the incremental filling of the cache as a user navigates the dataset along denormalized dimensions, which is a common pattern that the cache optimizes for.

LLAP caches metadata and data from the input files. To maintain the validity of the cache in the presence of file updates, LLAP uses a unique identifier assigned to every file stored in HDFS together with information about the file length. This is similar to the ETag fields present in blob stores such as AWS S3 or Azure Blob Storage.

The metadata, including index information, is cached even for data that was never in the cache. In particular, the first scan populates the metadata in bulk, which is used to determine the row groups that have to be loaded and to evaluate predicates before determining cache misses. The advantage of this approach is that LLAP will not load chunks that are effectively unnecessary for a given query, and hence will avoid trashing the cache.

The cache is incrementally mutable, i.e., the addition of new data to the table does not result in a complete cache invalidation. Specifically, the cache acquires a file only when a query addresses that file, which transfers the control of the visibility back to the query transactional state. Since ACID implementation handles transactions by adjusting visibility at the file level (recall Section~\ref{subsec:acid}), the cache turns into an MVCC view of the data servicing multiple concurrent queries possibly in different transactional states.

The eviction policy for the data in the cache is exchangeable. Currently a simple LRFU (Least Recently/Frequently Used) replacement policy that is tuned for analytic workloads with frequent full and partial scan operations is used by default. The unit of data for eviction is the chunk. This choice represents a compromise between low-overhead processing and storage efficiency.

\myparagraph{Query fragment execution.} LLAP daemons execute arbitrary query plan \textit{fragments} containing operations such as filters, projections, data transformations, joins, partial aggregates, and sorting. They allow parallel execution for multiple fragments from different queries and sessions. The operations in the fragments are vectorized and they run directly on the internal RLE format. Using the same format for I/O, cache, and execution minimizes the work needed for interaction among all of them.

As mentioned above, LLAP does not contain its own execution logic. Instead, its executors can essentially replicate YARN containers functionality. However, for stability and security reasons, only Hive code and statically specified user-defined functions (UDFs) are accepted in LLAP.

\mysubsection{Workload management improvements}
\label{subsec:wm}

The workload manager controls the access to LLAP resources for each query executed by Hive. An administrator can create \textit{resource plans}, i.e., self-contained resource-sharing configurations, to improve execution predictability and cluster sharing by concurrent queries running on LLAP. These factors are critical in multi-tenant environments. Though multiple resource plans can be defined in the system, at a given time only one of them can be active for a given deployment. Resource plans are persisted by Hive in HMS. 

A resource plan consists of ($i$)~one or more pool of resources, with a maximum amount of resources and number of concurrent queries per pool, ($ii$)~mappings, which root incoming queries to pools based on specific query properties, such as user, group, or application, and ($iii$)~triggers which initiate an action, such as killing queries in a pool or moving queries from one pool to another, based on query metrics that are collected at runtime. Though queries get guaranteed fractions of the cluster resources as defined in the pools, the workload manager tries to prevent that the cluster is underutilized. In particular, a query may be assigned idle resources from a pool that it has not been assigned to, until a subsequent query that maps to that pool claims them.

To provide an example, consider the following resource plan definition for a production cluster:
\begin{lstlisting}[style=HIVESQL]
CREATE RESOURCE PLAN daytime;
CREATE POOL daytime.bi
  WITH alloc_fraction=0.8, query_parallelism=5;
CREATE POOL daytime.etl
  WITH alloc_fraction=0.2, query_parallelism=20;
CREATE RULE downgrade IN daytime
  WHEN total_runtime > 3000 THEN MOVE etl;
ADD RULE downgrade TO bi;
CREATE APPLICATION MAPPING visualization_app IN daytime TO bi;
ALTER PLAN daytime SET DEFAULT POOL = etl;
ALTER RESOURCE PLAN daytime ENABLE ACTIVATE;
\end{lstlisting}

Line~$1$ creates the resource plan \textit{daytime}. Lines~$2$-$3$ create a pool \textit{bi} with $80\%$ of the LLAP resources in the cluster. Those resources may be used to execute up to $5$ queries concurrently. Similarly, lines~$4$-$5$ create a pool \textit{etl} with the rest of resources that may be used to execute up to $20$ queries concurrently. Lines $6$-$8$ create a rule that moves a query from the \textit{bi} to the \textit{etl} pool of resources when the query has run for more than $3$ seconds. Note that the previous operation can be executed because query fragments are easier to pre-empt compared to containers. Line $9$ creates a mapping for an application called \textit{interactive\_bi}, i.e., all queries fired by \textit{interactive\_bi} will initially take resources from the \textit{bi} pool. In turn, line $10$ sets the default pool to \textit{etl} for the rest of queries in the system. Finally, line $11$ enables and activates the resource plan in the cluster.

%%%%%SECTION%%%%%
\mysection{Federated warehouse system}
\label{sec:fed}

Over the last decade, there has been a growing proliferation of specialized data management systems~\cite{DBLP:conf/icde/StonebrakerC05} which have become popular because they achieve better cost-effective performance for their specific use case than traditional RDBMSs.

In addition to its native processing capabilities, Hive can act as a \textit{mediator}~\cite{DBLP:journals/computer/Wiederhold92,DBLP:conf/ride/CareyHSACFFLNPTWW95,DBLP:conf/cidr/BugiottiBDIM15} as it is designed to support querying over multiple independent data management systems. The benefits of unifying access to these systems through Hive are multifold. Application developers can choose a blend of multiple systems to achieve the desired performance and functionality, yet they need to code only against a single interface. Thus, applications become independent of the underlying data systems, which allows for more flexibility in changing systems later. Hive itself can be used to implement data movement and transformations between different systems, alleviating the need for third-party tools. Besides, Hive as a mediator can globally enforce access control and capture audit trails via Ranger~\cite{website:Ranger} or Sentry~\cite{website:Sentry}, and also help with compliance requirements via Atlas~\cite{website:Atlas}.

\mysubsection{Storage handlers}
To interact with other engines in a modular and extensible fashion, Hive includes a \textit{storage handler} interface that needs to be implemented for each of them. A storage handler consists of: ($i$)~an \textit{input format}, which describes how to read data from the external engine, including how to split the work to increase parallelism, ($ii$)~an \textit{output format}, which describes how to write data to the external engine, ($iii$)~a \textit{SerDe} (serializer and deserializer) which describes how to transform data from Hive internal representation into the external engine representation and vice versa, and ($iv$)~a \textit{Metastore hook}, which defines notification methods invoked as part of the transactions against HMS, e.g., when a new table backed by the external system is created or when new rows are inserted into such table. The minimum implementation of a usable storage handler to read data from the external system contains at least an input format and deserializer.

Once the storage handler interface has been implemented, querying the external system from Hive is straightforward and all the complexity is hidden from user behind the storage handler implementation. For instance, Apache Druid~\cite{DBLP:conf/sigmod/YangTLRMG14,website:Druid} is an open-source data store designed for business intelligence (OLAP) queries on event data that is widely used to power user-facing analytic applications; Hive provides a Druid storage handler so it can take advantage of its efficiency for the execution of interactive queries. To start querying Druid from Hive, the only action necessary is to register or create Druid data sources from Hive. First, if a data source already exists in Druid, we can map a Hive external table to it with a simple statement:

\begin{lstlisting}[style=HIVESQL]
CREATE EXTERNAL TABLE druid_table_1
STORED BY 'org.apache.hadoop.hive.druid.DruidStorageHandler'
TBLPROPERTIES ('druid.datasource' = 'my_druid_source');
\end{lstlisting}

Observe that we do not need to specify column names or types for the data source, since they are automatically inferred from Druid metadata. In turn, we can create a data source in Druid from Hive with a simple statement as follows:

\begin{lstlisting}[style=HIVESQL]
CREATE EXTERNAL TABLE druid_table_2 (
  __time TIMESTAMP, dim1 VARCHAR(20), m1 FLOAT)
STORED BY 'org.apache.hadoop.hive.druid.DruidStorageHandler';
\end{lstlisting}

Once the Druid sources are available in Hive as external tables, we can execute any allowed table operations on them.

\mysubsection{Pushing computation using Calcite}
One of the most powerful features of Hive is the possibility to leverage Calcite adapters~\cite{DBLP:conf/sigmod/BegoliCHML18} to push complex computation to supported systems and generate queries in the languages supported by those systems.

Continuing with the Druid example, the most common way of querying Druid is through a REST API over HTTP using queries expressed in JSON~\footnote{\url{http://druid.io/docs/latest/querying/querying.html}}. Once a user has declared a table that is stored in Druid, Hive can transparently generate Druid JSON queries from the input SQL queries. In particular, the optimizer applies rules that match a sequence of operators in the plan and generate a new equivalent sequence with more operations executed in Druid. Once we have completed the optimization phase, the subset of operators that needs to be executed by Druid is translated by Calcite into a valid JSON query that is attached to the scan operator that will read from Druid. Note that for a storage handler to support Calcite automatically generated queries, its input format needs to include logic to send the query to the external system (possibly splitting the query into multiple sub-queries that can be executed in parallel) and read back the query results.

\begin{figure}[t]
\centering{
\setlength\extrarowheight{4pt}
\begin{tabular}{c c}
{\begin{lstlisting}[style=HIVESQL,numbers=none,xleftmargin=0pt]
SELECT d1, SUM(m1) AS s
FROM druid_table_1
WHERE EXTRACT(year FROM __time)
  BETWEEN 2017 AND 2018
GROUP BY d1
ORDER BY s DESC
LIMIT 10;
\end{lstlisting}} &
\raisebox{-.5\height}{\includegraphics[width=0.4\columnwidth]{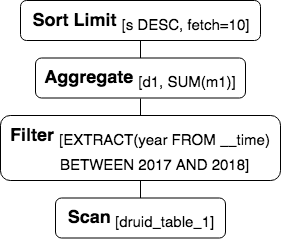}} \\[2ex]
\small{\textbf{(a)} SQL query over Druid table} & \small{\textbf{(b)} Plan before optimization} \\
\end{tabular}
\newline
\vspace*{6pt}
\newline
\begin{tabular}{c}
\includegraphics[width=0.3\columnwidth]{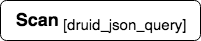}\\
{\begin{lstlisting}[style=JSON,numbers=none,xleftmargin=10pt,frameshape={ryrRYR}{yY}{yY}{ryrRYR},linewidth=0.905\columnwidth]
{
  "queryType": "groupBy",
  "dataSource": "my_druid_source",
  "granularity": "all",
  "dimension": "d1",
  "aggregations":
    [ { "type": "floatSum", "name": "s", "fieldName": "m1" } ],
  "limitSpec": {
    "limit": 10,
    "columns":
      [ {"dimension": "s", "direction": "descending" } ]
  },
  "intervals":
    [ "2017-01-01T00:00:00.000/2019-01-01T00:00:00.000" ]
}
\end{lstlisting}} \\
\small{\textbf{(c)} Plan after optimization} \\
\end{tabular}}
\vspace{-2mm}\caption{Query federation example in Hive.\label{fig:federation}}\vspace{-3mm}
\end{figure}

As of today, Hive can push operations to Druid and multiple engines with JDBC support\footnote{Calcite can generate SQL queries from operator expressions using a large number of different dialects.} using Calcite. Figure~\ref{fig:federation} depicts a query executed over a table stored in Druid, and the corresponding plan and JSON query generated by Calcite.

%\input{repl}
%%%%%SECTION%%%%%
\mysection{Performance evaluation}
\label{sec:exp}

To study the impact of our work, we evaluate Hive over time using different standard benchmarks. In this section, we present a summary of results that highlight the impact of the improvements presented throughout this paper.

\myparagraph{Experimental setup.} The experiments presented below were run on a cluster of $10$ nodes connected by a $10$ Gigabit Ethernet network. Each node has a $8$-core $2.40$GHz Intel Xeon CPU E$5$-$2630$ v$3$ and $256$GB RAM, and has two $6$TB disks for HDFS and YARN storage.

\begin{figure*}[t]
\centering
\includegraphics[trim={1.75cm 0 0.5cm 0},width=0.99\textwidth,clip]{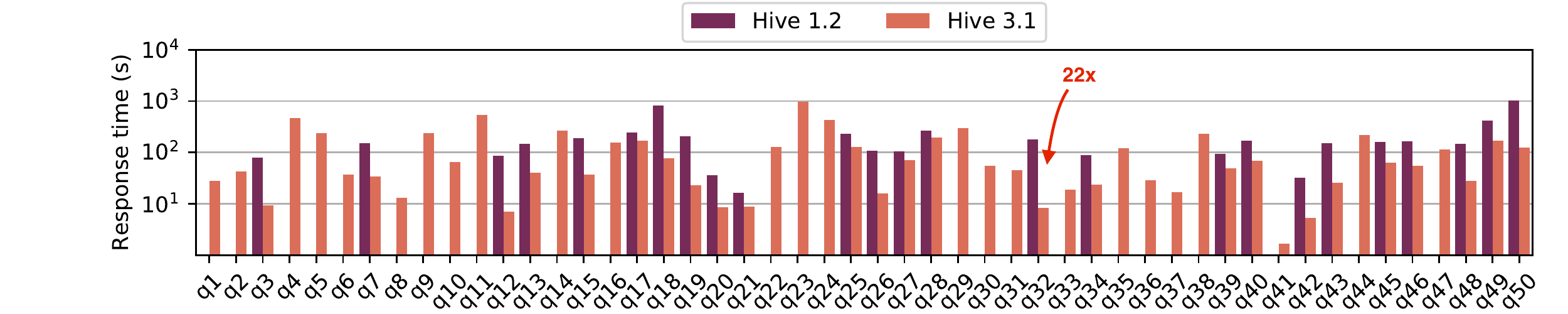}
\includegraphics[trim={1.75cm 0 0.5cm 0.60cm},width=0.99\textwidth,clip]{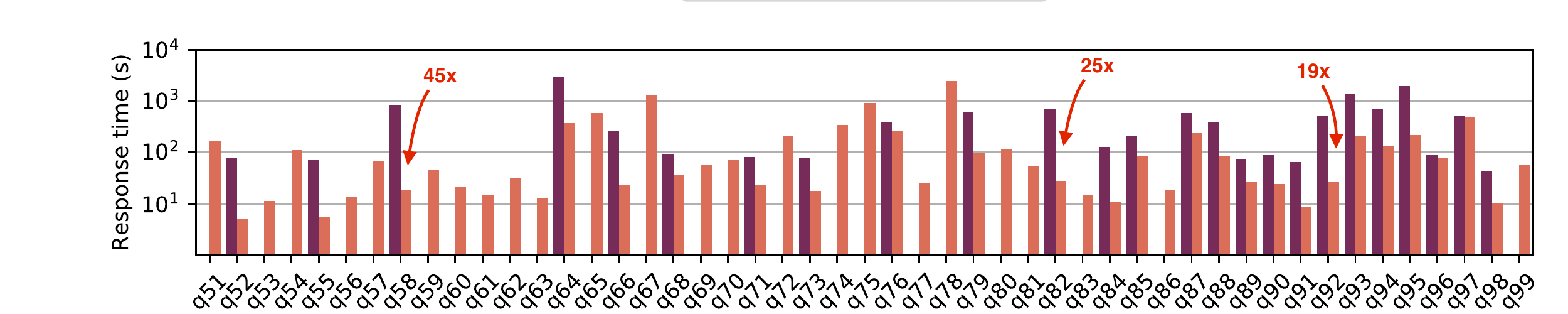}
\vspace{-2mm}\caption{Comparison of query response times among different Hive versions.\label{fig:tpcds}}
\end{figure*}

\mysubsection{Comparison with previous versions}

We have conducted experiments using TPC-DS queries on a $10$TB scale data set. The data was stored in ACID tables in HDFS using the ORC file format, and the fact tables were partitioned by day. All the information needed for reproducibility of these results is publicly available~\footnote{\url{http://github.com/hortonworks/hive-testbench/tree/hdp3}}. We compared two different versions of Hive~\cite{website:Hive} with each other: ($i$)~Hive v$1.2$, released in September 2015, running on top of Tez v$0.5$, and ($ii$)~the latest Hive v$3.1$, released in November 2018, using Tez v$0.9$ with LLAP enabled. Figure~\ref{fig:tpcds} shows the response times (perceived by the user) for both Hive versions; note the logarithmic $y$ axis. For each query, we report the average over three runs with warm cache. First, observe that only $50$ queries could be executed in Hive v$1.2$; the response time values for the queries that could not be executed are omitted in the figure. The reason is that Hive v$1.2$ lacked support for set operations such as \texttt{EXCEPT} or \texttt{INTERSECT}, correlated scalar subqueries with non-equi join conditions, interval notation, and order by unselected columns, among other SQL features. For those $50$ queries, Hive v$3.1$ significantly outperforms the preceding version. In particular, Hive v$3.1$ is faster by an average of $4.6$x and by up to a maximum of $45.5$x (refer to \textit{q58}). Queries that improved more than $15$x are emphasized in the figure. More importantly, Hive v$3.1$ can execute the full set of $99$ TPC-DS queries thanks to its SQL support improvements. The performance difference between both versions is so significant that the aggregated response time for all queries executed by Hive v$3.1$ is still $15\%$ lower than the time for $50$ queries in Hive v$1.2$. New optimization features such as shared work optimizer make a big difference on their own; for example, \textit{q88} is $2.7$x faster when it is enabled.

\mysubsection{LLAP acceleration}
In order to illustrate the benefits of LLAP over query execution using Tez containers uniquely, we run all the $99$ TPC-DS queries in Hive v$3.1$ using the same configuration but with LLAP enabled/disabled. Table~\ref{tab:llap} shows the aggregated time for all the queries in the experiment; we can observe that LLAP on its own reduces the workload response time dramatically by $2.7$x.

\begin{table}[t]
\begin{tabular}{|c|r|}
\hline
\textbf{Execution mode} & \textbf{Total response time (s)} \\
\hline\hline
Container (without LLAP) & $41576$ \\
\hline
LLAP & $15540$\\
\hline
\end{tabular}
\caption{Response time improvement using LLAP.\label{tab:llap}}\vspace{-3mm}
\end{table}

\begin{figure}[t]
\centering
\vspace{-0.2cm}\includegraphics[width=0.99\columnwidth]{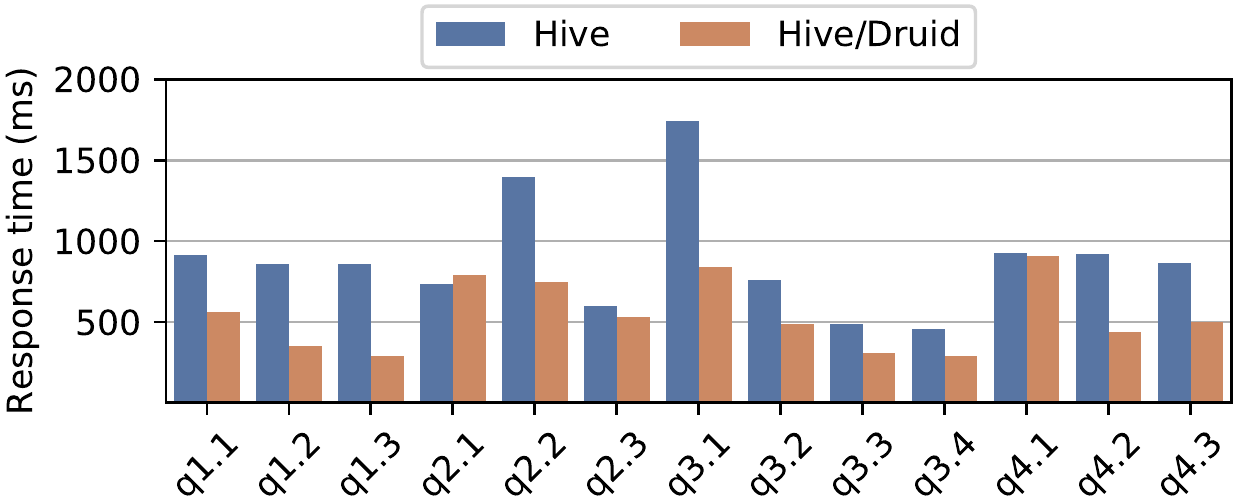}
\vspace{-0.2cm}\caption{Comparison of query response times between native Hive and federation to Druid.\label{fig:ssb}}\vspace{-0.1cm}
\end{figure}

\mysubsection{Query federation to Druid}

In the following, we illustrate the significant performance benefits that can be derived from the combination of using Hive's materialized views and federation capabilities. For this experiment, we use the Star-Schema Benchmark (SSB)~\cite{DBLP:conf/tpctc/ONeilOCR09} on a 1TB scale data set. The SSB benchmark is based on TPC-H and it is meant to simulate the process of iteratively and interactively querying a data warehouse to play what-if scenarios, drill down and better understand trends. It consists of a single fact table and $4$ dimension tables, and the workload contains $13$ queries that join, aggregate, and place fairly tight dimensional filters over different sets of tables.

For the experiment, we create a materialized view that denormalizes the database schema~\footnote{\url{https://hortonworks.com/blog/sub-second-analytics-hive-druid/}}. The materialization is stored in Hive. Then we run the queries in the benchmark which are automatically rewritten by the optimizer to be answered from the materialized view. Subsequently, we store the materialized view in Druid v$0.12$ and repeat the same steps. Figure~\ref{fig:ssb} depicts response times for each variant. Observe that Hive/Druid is $1.6$x faster than execution over the materialized view stored natively in Hive. The reason is that Hive pushes most of the query computation to Druid using Calcite, and thus it can benefit from Druid providing lower latency for those queries.

%It would be interesting to mention the size of the preaggregated cube. We should include it in the CR.

%%%%%SECTION%%%%%
\mysection{Discussion}
\label{sec:discussion}

In this section, we discuss the impact of the improvements presented in this work, and multiple challenges faced and lessons learned while implementing these new features.

The decision on the features that should be added to the project over time is mainly influenced by shortcomings faced by users, either reported directly to the Hive community or through the companies commercializing the project.

For instance, enterprise users asked for the implementation of ACID guarantees in order to offload workloads from data warehouses introduced in their organizations long before Hadoop. Besides, new regulations coming into effect recently, such as the European GDPR that gives individuals the right to request erasure of personal data, have only emphasized the importance of this new feature. Not surprisingly, ACID support has quickly become a key differentiator to choose Hive over other SQL engines on top of Hadoop. However, bringing ACID capabilities to Hive was not straightforward and the initial implementation had to go through major changes because it introduced a reading latency penalty, compared to non-ACID tables, that was unacceptable for users. This was due to a detail in the original design that we did not foresee having such a large impact on performance in production: Using a single delta type of file to store inserted, updated, and deleted records. If a workload consisted of a large number of writing operations or compaction was not run frequently enough, the file readers had to perform a sort-merge across a large number of base and delta files to consolidate the data, potentially creating memory pressure. Further, filter pushdown to skip reading entire row groups could not be applied to those delta files. The second implementation of ACID, described in this paper, was rolled out with Hive v$3.1$. It solves the issues described above and performance is at par with non-ACID tables.

Another common and frequent request by Hive users over time was to improve its runtime latency. LLAP development started $4$~years ago to address issues that were inherent to the architecture of the system, such as the lack of data caching or long running executors. The feedback from initial deployments was quickly incorporated to the system. For instance, predictability was a huge concern in production due to cluster resource contention among user queries, which lead to the implementation of the workload manager. Currently it is extremely rewarding to see companies deploying LLAP to serve queries over TBs of data in multi-tenant clusters, achieving average latency in the order of seconds~\footnote{\url{https://azure.microsoft.com/en-us/blog/hdinsight-interactive-query-performance-benchmarks-and-integration-with-power-bi-direct-query/}}.

Another important decision was the integration with Calcite for query optimization. Representing queries at the right abstraction level was critical to implementing advanced optimization algorithms in Hive, which in turn brought huge performance benefits to the system. Besides, the integration was leveraged to easily generate queries for federation to other systems. Since nowadays it is common for organizations to use a plethora of data management systems, this new feature has been received with lots of excitement by Hive users.

%%%%%SECTION%%%%%
\mysection{Roadmap}
\label{sec:roadmap}

\myparagraph{Continuous improvements to Hive.} The community will continue working on the areas discussed in this paper to make Hive a better warehousing solution. For instance, based on the work that we have already completed to support ACID capabilities, we plan to implement multi-statement transactions. In addition, we shall continue improving the optimization techniques used in Hive. Statistics captured at runtime for query reoptimization are already persisted in HMS, which will allow us to feedback that information into the optimizer to obtain more accurate estimates, similar to other systems~\cite{DBLP:conf/vldb/StillgerLMK01,DBLP:journals/pvldb/ZaitCBVKW17}. Materialized views work is still ongoing and one of the most requested features is the implementation of an advisor or recommender~\cite{DBLP:conf/vldb/AgrawalCN00,DBLP:conf/vldb/ZilioRLLSGF04} for Hive. Improvements to LLAP performance and stability as well as the implementation of new connectors to other specialized systems, e.g., Kafka~\cite{website:Kafka,DBLP:Kreps2011Kafka}, are in progress too.

\myparagraph{Standalone Metastore.} HMS has become a critical part in Hadoop as it is used by other data processing engines such as Spark or Presto to provide a central repository with information about all data sources in each system. Hence, there is a growing interest and ongoing work to spin-off the Metastore from Hive and to develop it as a standalone project. Each of these systems will have their own catalog in the Metastore and it will be easier to access data across different engines.

\myparagraph{Containerized Hive in the cloud.} Cloud-based data warehousing solutions such as Azure SQL DW~\cite{website:AzureSQLDW}, Redshift~\cite{website:Redshift,DBLP:conf/sigmod/GuptaATKPSS15}, BigQuery~\cite{website:BigQuery,DBLP:journals/pvldb/MelnikGLRSTV10} and Snowflake~\cite{website:Snowflake,DBLP:conf/sigmod/DagevilleCZAABC16} have been gaining popularity over the last few years. Besides, there is an increasing interest in systems such as Kubernetes~\cite{website:Kubernetes} to provide container orchestration for applications deployed indistinctly on-premises or in the cloud. Hive's modular architecture makes it easy to isolate its components (HS2, HMS, LLAP) and run them in containers. Ongoing effort is focused on finalizing the work to enable the deployment of Hive in commercial clouds using Kubernetes.

%%%%%SECTION%%%%%
\mysection{Conclusions}
\label{sec:conclusion}
Apache Hive's early success stemmed from the ability to exploit parallelism for batch operations with a well-known interface. It made data load and management simple, handling node, software and hardware failures gracefully without expensive repair or recovery times.

In this paper we show how the community expanded the utility of the system from ETL tool to fully-fledged enterprise-grade data warehouse. We described the addition of a transactional system that is well suited for the data modifications required in star schema databases. We showed the main runtime improvements necessary to bring query latency and concurrency into the realm of interactive operation. We also described cost-based optimization techniques necessary to handle today's view hierarchies and big-data operations. Finally, we showed how Hive can be used today as a relational front-end to multiple storage and data systems. All of this happened without ever compromising on the original characteristics of the system that made it popular.
%SQL is still the standard interface and the available dialect has grown dramatically, and data load and management remains simple and is possible on a myriad of storage systems on-premises and in the cloud. Regardless of execution mode, whether MapReduce, Tez or LLAP, \sys{} handles node, software and hardware failures gracefully without expensive repair or recovery times.

Apache Hive architecture and design principles have proven to be powerful in today's analytic landscape. We believe it will continue to thrive in new deployment and storage environments as they emerge, as it is showing today with containerization and cloud.

\begin{acks}
We would like to thank the Apache Hive community, contributors and users, who build, maintain, use, test, write about, and continue to push the project forward. 
We would also like to thank Oliver Draese for his feedback on the original draft of this paper.
\end{acks}

\balance

\bibliographystyle{ACM-Reference-Format}
\bibliography{references}


\begin{thebibliography}{61}

%%% the Web address.

\ifx \showCODEN    \undefined \def \showCODEN     #1{\unskip}     \fi
\ifx \showDOI      \undefined \def \showDOI       #1{#1}\fi
\ifx \showISBNx    \undefined \def \showISBNx     #1{\unskip}     \fi
\ifx \showISBNxiii \undefined \def \showISBNxiii  #1{\unskip}     \fi
\ifx \showISSN     \undefined \def \showISSN      #1{\unskip}     \fi
\ifx \showLCCN     \undefined \def \showLCCN      #1{\unskip}     \fi
\ifx \shownote     \undefined \def \shownote      #1{#1}          \fi
\ifx \showarticletitle \undefined \def \showarticletitle #1{#1}   \fi
\ifx \showURL      \undefined \def \showURL       {\relax}        \fi
% invisible to TeX
\providecommand\bibfield[2]{#2}
\providecommand\bibinfo[2]{#2}
\providecommand\natexlab[1]{#1}
\providecommand\showeprint[2][]{arXiv:#2}

\bibitem[\protect\citeauthoryear{Agrawal, Chaudhuri, and Narasayya}{Agrawal
  et~al\mbox{.}}{2000}]%
        {DBLP:conf/vldb/AgrawalCN00}
\bibfield{author}{\bibinfo{person}{Sanjay Agrawal}, \bibinfo{person}{Surajit
  Chaudhuri}, {and} \bibinfo{person}{Vivek~R. Narasayya}.}
  \bibinfo{year}{2000}\natexlab{}.
\newblock \showarticletitle{Automated Selection of Materialized Views and
  Indexes in {SQL} Databases}. In \bibinfo{booktitle}{\emph{{PVLDB}}}.
\newblock


\bibitem[\protect\citeauthoryear{Apache Atlas}{Apache Atlas}{2018}]%
        {website:Atlas}
Apache Atlas \bibinfo{year}{2018}\natexlab{}.
\newblock \bibinfo{title}{{Apache Atlas}: Data Governance and Metadata
  framework for Hadoop}.
\newblock
\newblock
\newblock
\shownote{http://atlas.apache.org/.}


\bibitem[\protect\citeauthoryear{Apache Calcite}{Apache Calcite}{2018}]%
        {website:Calcite}
Apache Calcite \bibinfo{year}{2018}\natexlab{}.
\newblock \bibinfo{title}{{Apache Calcite}: Dynamic data management framework}.
\newblock
\newblock
\newblock
\shownote{http://calcite.apache.org/.}


\bibitem[\protect\citeauthoryear{Apache Druid}{Apache Druid}{2018}]%
        {website:Druid}
Apache Druid \bibinfo{year}{2018}\natexlab{}.
\newblock \bibinfo{title}{{Apache Druid}: Interactive analytics at scale}.
\newblock
\newblock
\newblock
\shownote{http://druid.io/.}


\bibitem[\protect\citeauthoryear{Apache Flink}{Apache Flink}{2018}]%
        {website:Flink}
Apache Flink \bibinfo{year}{2018}\natexlab{}.
\newblock \bibinfo{title}{{Apache Flink}: Stateful Computations over Data
  Streams}.
\newblock
\newblock
\newblock
\shownote{http://flink.apache.org/.}


\bibitem[\protect\citeauthoryear{Apache HBase}{Apache HBase}{2018}]%
        {website:HBase}
Apache HBase \bibinfo{year}{2018}\natexlab{}.
\newblock \bibinfo{title}{{Apache HBase}}.
\newblock
\newblock
\newblock
\shownote{http://hbase.apache.org/.}


\bibitem[\protect\citeauthoryear{Apache Hive}{Apache Hive}{2018}]%
        {website:Hive}
Apache Hive \bibinfo{year}{2018}\natexlab{}.
\newblock \bibinfo{title}{{Apache Hive}}.
\newblock
\newblock
\newblock
\shownote{http://hive.apache.org/.}


\bibitem[\protect\citeauthoryear{Apache Impala}{Apache Impala}{2018}]%
        {website:Impala}
Apache Impala \bibinfo{year}{2018}\natexlab{}.
\newblock \bibinfo{title}{{Apache Impala}}.
\newblock
\newblock
\newblock
\shownote{http://impala.apache.org/.}


\bibitem[\protect\citeauthoryear{Apache Kafka}{Apache Kafka}{2018}]%
        {website:Kafka}
Apache Kafka \bibinfo{year}{2018}\natexlab{}.
\newblock \bibinfo{title}{{Apache Kafka}: A distributed streaming platform}.
\newblock
\newblock
\newblock
\shownote{http://kafka.apache.org/.}


\bibitem[\protect\citeauthoryear{Apache ORC}{Apache ORC}{2018}]%
        {website:Orc}
Apache ORC \bibinfo{year}{2018}\natexlab{}.
\newblock \bibinfo{title}{{Apache ORC}: High-Performance Columnar Storage for
  Hadoop}.
\newblock
\newblock
\newblock
\shownote{http://orc.apache.org/.}


\bibitem[\protect\citeauthoryear{Apache Parquet}{Apache Parquet}{2018}]%
        {website:Parquet}
Apache Parquet \bibinfo{year}{2018}\natexlab{}.
\newblock \bibinfo{title}{{Apache Parquet}}.
\newblock
\newblock
\newblock
\shownote{http://parquet.apache.org/.}


\bibitem[\protect\citeauthoryear{Apache Ranger}{Apache Ranger}{2018}]%
        {website:Ranger}
Apache Ranger \bibinfo{year}{2018}\natexlab{}.
\newblock \bibinfo{title}{{Apache Ranger}: Framework to enable, monitor and
  manage comprehensive data security across the Hadoop platform.}
\newblock
\newblock
\newblock
\shownote{http://ranger.apache.org/.}


\bibitem[\protect\citeauthoryear{Apache Sentry}{Apache Sentry}{2018}]%
        {website:Sentry}
Apache Sentry \bibinfo{year}{2018}\natexlab{}.
\newblock \bibinfo{title}{{Apache Sentry}: System for enforcing fine grained
  role based authorization to data and metadata stored on a Hadoop cluster.}
\newblock
\newblock
\newblock
\shownote{http://sentry.apache.org/.}


\bibitem[\protect\citeauthoryear{Apache Spark}{Apache Spark}{2018}]%
        {website:Spark}
Apache Spark \bibinfo{year}{2018}\natexlab{}.
\newblock \bibinfo{title}{{Apache Spark}: Unified Analytics Engine for Big
  Data}.
\newblock
\newblock
\newblock
\shownote{http://spark.apache.org/.}


\bibitem[\protect\citeauthoryear{Apache Tez}{Apache Tez}{2018}]%
        {website:Tez}
Apache Tez \bibinfo{year}{2018}\natexlab{}.
\newblock \bibinfo{title}{{Apache Tez}}.
\newblock
\newblock
\newblock
\shownote{http://tez.apache.org/.}


\bibitem[\protect\citeauthoryear{Apache Thrift}{Apache Thrift}{2018}]%
        {website:Thrift}
Apache Thrift \bibinfo{year}{2018}\natexlab{}.
\newblock \bibinfo{title}{{Apache Thrift}}.
\newblock
\newblock
\newblock
\shownote{http://thrift.apache.org/.}


\bibitem[\protect\citeauthoryear{Apers, Hevner, and Yao}{Apers
  et~al\mbox{.}}{1983}]%
        {DBLP:journals/tse/ApersHY83}
\bibfield{author}{\bibinfo{person}{Peter M.~G. Apers}, \bibinfo{person}{Alan~R.
  Hevner}, {and} \bibinfo{person}{S.~Bing Yao}.}
  \bibinfo{year}{1983}\natexlab{}.
\newblock \showarticletitle{Optimization Algorithms for Distributed Queries}.
\newblock \bibinfo{journal}{\emph{{IEEE} Trans. Software Eng.}}
  \bibinfo{volume}{9}, \bibinfo{number}{1} (\bibinfo{year}{1983}),
  \bibinfo{pages}{57--68}.
\newblock


\bibitem[\protect\citeauthoryear{Azure SQL DW}{Azure SQL DW}{2018}]%
        {website:AzureSQLDW}
Azure SQL DW \bibinfo{year}{2018}\natexlab{}.
\newblock \bibinfo{title}{{Azure SQL Data Warehouse}}.
\newblock
\newblock
\newblock
\shownote{http://azure.microsoft.com/en-us/services/sql-data-warehouse/.}


\bibitem[\protect\citeauthoryear{Begoli, Camacho{-}Rodr{\'{\i}}guez, Hyde,
  Mior, and Lemire}{Begoli et~al\mbox{.}}{2018}]%
        {DBLP:conf/sigmod/BegoliCHML18}
\bibfield{author}{\bibinfo{person}{Edmon Begoli}, \bibinfo{person}{Jes{\'{u}}s
  Camacho{-}Rodr{\'{\i}}guez}, \bibinfo{person}{Julian Hyde},
  \bibinfo{person}{Michael~J. Mior}, {and} \bibinfo{person}{Daniel Lemire}.}
  \bibinfo{year}{2018}\natexlab{}.
\newblock \showarticletitle{Apache Calcite: {A} Foundational Framework for
  Optimized Query Processing Over Heterogeneous Data Sources}. In
  \bibinfo{booktitle}{\emph{{SIGMOD}}}.
\newblock


\bibitem[\protect\citeauthoryear{Bernstein, Goodman, Wong, Reeve, and
  Jr.}{Bernstein et~al\mbox{.}}{1981}]%
        {DBLP:journals/tods/BernsteinGWRR81}
\bibfield{author}{\bibinfo{person}{Philip~A. Bernstein},
  \bibinfo{person}{Nathan Goodman}, \bibinfo{person}{Eugene Wong},
  \bibinfo{person}{Christopher~L. Reeve}, {and} \bibinfo{person}{James
  B.~Rothnie Jr.}} \bibinfo{year}{1981}\natexlab{}.
\newblock \showarticletitle{Query Processing in a System for Distributed
  Databases {(SDD-1)}}.
\newblock \bibinfo{journal}{\emph{{ACM} Trans. Database Syst.}}
  \bibinfo{volume}{6}, \bibinfo{number}{4} (\bibinfo{year}{1981}),
  \bibinfo{pages}{602--625}.
\newblock


\bibitem[\protect\citeauthoryear{BigQuery}{BigQuery}{2018}]%
        {website:BigQuery}
BigQuery \bibinfo{year}{2018}\natexlab{}.
\newblock \bibinfo{title}{{BigQuery}: Analytics Data Warehouse}.
\newblock
\newblock
\newblock
\shownote{http://cloud.google.com/bigquery/.}


\bibitem[\protect\citeauthoryear{Borkar, Carey, Grover, Onose, and
  Vernica}{Borkar et~al\mbox{.}}{2011}]%
        {DBLP:conf/icde/BorkarCGOV11}
\bibfield{author}{\bibinfo{person}{Vinayak~R. Borkar},
  \bibinfo{person}{Michael~J. Carey}, \bibinfo{person}{Raman Grover},
  \bibinfo{person}{Nicola Onose}, {and} \bibinfo{person}{Rares Vernica}.}
  \bibinfo{year}{2011}\natexlab{}.
\newblock \showarticletitle{Hyracks: {A} flexible and extensible foundation for
  data-intensive computing}. In \bibinfo{booktitle}{\emph{{ICDE}}}.
\newblock


\bibitem[\protect\citeauthoryear{Bugiotti, Bursztyn, Deutsch, Ileana, and
  Manolescu}{Bugiotti et~al\mbox{.}}{2015}]%
        {DBLP:conf/cidr/BugiottiBDIM15}
\bibfield{author}{\bibinfo{person}{Francesca Bugiotti}, \bibinfo{person}{Damian
  Bursztyn}, \bibinfo{person}{Alin Deutsch}, \bibinfo{person}{Ioana Ileana},
  {and} \bibinfo{person}{Ioana Manolescu}.} \bibinfo{year}{2015}\natexlab{}.
\newblock \showarticletitle{Invisible Glue: Scalable Self-Tunning
  Multi-Stores}. In \bibinfo{booktitle}{\emph{{CIDR}}}.
\newblock


\bibitem[\protect\citeauthoryear{Cahill, R{\"{o}}hm, and Fekete}{Cahill
  et~al\mbox{.}}{2008}]%
        {DBLP:conf/sigmod/CahillRF08}
\bibfield{author}{\bibinfo{person}{Michael~J. Cahill}, \bibinfo{person}{Uwe
  R{\"{o}}hm}, {and} \bibinfo{person}{Alan~David Fekete}.}
  \bibinfo{year}{2008}\natexlab{}.
\newblock \showarticletitle{Serializable isolation for snapshot databases}. In
  \bibinfo{booktitle}{\emph{{SIGMOD}}}.
\newblock


\bibitem[\protect\citeauthoryear{Camacho{-}Rodr{\'{\i}}guez, Colazzo, Herschel,
  Manolescu, and Chowdhury}{Camacho{-}Rodr{\'{\i}}guez et~al\mbox{.}}{2016}]%
        {DBLP:conf/cikm/Camacho-Rodriguez16}
\bibfield{author}{\bibinfo{person}{Jes{\'{u}}s Camacho{-}Rodr{\'{\i}}guez},
  \bibinfo{person}{Dario Colazzo}, \bibinfo{person}{Melanie Herschel},
  \bibinfo{person}{Ioana Manolescu}, {and} \bibinfo{person}{Soudip~Roy
  Chowdhury}.} \bibinfo{year}{2016}\natexlab{}.
\newblock \showarticletitle{Reuse-based Optimization for Pig Latin}. In
  \bibinfo{booktitle}{\emph{{CIKM}}}.
\newblock


\bibitem[\protect\citeauthoryear{Carbone, Katsifodimos, Ewen, Markl, Haridi,
  and Tzoumas}{Carbone et~al\mbox{.}}{2015}]%
        {DBLP:journals/debu/CarboneKEMHT15}
\bibfield{author}{\bibinfo{person}{Paris Carbone}, \bibinfo{person}{Asterios
  Katsifodimos}, \bibinfo{person}{Stephan Ewen}, \bibinfo{person}{Volker
  Markl}, \bibinfo{person}{Seif Haridi}, {and} \bibinfo{person}{Kostas
  Tzoumas}.} \bibinfo{year}{2015}\natexlab{}.
\newblock \showarticletitle{Apache Flink{\texttrademark}: Stream and Batch
  Processing in a Single Engine}.
\newblock \bibinfo{journal}{\emph{{IEEE} Data Eng. Bull.}}
  \bibinfo{volume}{38}, \bibinfo{number}{4} (\bibinfo{year}{2015}),
  \bibinfo{pages}{28--38}.
\newblock


\bibitem[\protect\citeauthoryear{Carey, Haas, Schwarz, Arya, Cody, Fagin,
  Flickner, Luniewski, Niblack, Petkovic, Thomas, Williams, and Wimmers}{Carey
  et~al\mbox{.}}{1995}]%
        {DBLP:conf/ride/CareyHSACFFLNPTWW95}
\bibfield{author}{\bibinfo{person}{Michael~J. Carey}, \bibinfo{person}{Laura~M.
  Haas}, \bibinfo{person}{Peter~M. Schwarz}, \bibinfo{person}{Manish Arya},
  \bibinfo{person}{William~F. Cody}, \bibinfo{person}{Ronald Fagin},
  \bibinfo{person}{Myron Flickner}, \bibinfo{person}{Allen Luniewski},
  \bibinfo{person}{Wayne Niblack}, \bibinfo{person}{Dragutin Petkovic},
  \bibinfo{person}{Joachim Thomas}, \bibinfo{person}{John~H. Williams}, {and}
  \bibinfo{person}{Edward~L. Wimmers}.} \bibinfo{year}{1995}\natexlab{}.
\newblock \showarticletitle{Towards Heterogeneous Multimedia Information
  Systems: The Garlic Approach}. In \bibinfo{booktitle}{\emph{{RIDE-DOM}
  Workshop}}.
\newblock


\bibitem[\protect\citeauthoryear{Chaudhuri, Krishnamurthy, Potamianos, and
  Shim}{Chaudhuri et~al\mbox{.}}{1995}]%
        {DBLP:conf/icde/ChaudhuriKPS95}
\bibfield{author}{\bibinfo{person}{Surajit Chaudhuri}, \bibinfo{person}{Ravi
  Krishnamurthy}, \bibinfo{person}{Spyros Potamianos}, {and}
  \bibinfo{person}{Kyuseok Shim}.} \bibinfo{year}{1995}\natexlab{}.
\newblock \showarticletitle{Optimizing Queries with Materialized Views}. In
  \bibinfo{booktitle}{\emph{{ICDE}}}.
\newblock


\bibitem[\protect\citeauthoryear{Dageville, Cruanes, Zukowski, Antonov, Avanes,
  Bock, Claybaugh, Engovatov, Hentschel, Huang, Lee, Motivala, Munir, Pelley,
  Povinec, Rahn, Triantafyllis, and Unterbrunner}{Dageville
  et~al\mbox{.}}{2016}]%
        {DBLP:conf/sigmod/DagevilleCZAABC16}
\bibfield{author}{\bibinfo{person}{Beno{\^{\i}}t Dageville},
  \bibinfo{person}{Thierry Cruanes}, \bibinfo{person}{Marcin Zukowski},
  \bibinfo{person}{Vadim Antonov}, \bibinfo{person}{Artin Avanes},
  \bibinfo{person}{Jon Bock}, \bibinfo{person}{Jonathan Claybaugh},
  \bibinfo{person}{Daniel Engovatov}, \bibinfo{person}{Martin Hentschel},
  \bibinfo{person}{Jiansheng Huang}, \bibinfo{person}{Allison~W. Lee},
  \bibinfo{person}{Ashish Motivala}, \bibinfo{person}{Abdul~Q. Munir},
  \bibinfo{person}{Steven Pelley}, \bibinfo{person}{Peter Povinec},
  \bibinfo{person}{Greg Rahn}, \bibinfo{person}{Spyridon Triantafyllis}, {and}
  \bibinfo{person}{Philipp Unterbrunner}.} \bibinfo{year}{2016}\natexlab{}.
\newblock \showarticletitle{The Snowflake Elastic Data Warehouse}. In
  \bibinfo{booktitle}{\emph{{SIGMOD}}}.
\newblock


\bibitem[\protect\citeauthoryear{DataNucleus}{DataNucleus}{2018}]%
        {website:DataNucleus}
DataNucleus \bibinfo{year}{2018}\natexlab{}.
\newblock \bibinfo{title}{{DataNucleus}: JDO/JPA/REST Persistence of Java
  Objects}.
\newblock
\newblock
\newblock
\shownote{http://www.datanucleus.org/.}


\bibitem[\protect\citeauthoryear{Goldstein and Larson}{Goldstein and
  Larson}{2001}]%
        {DBLP:conf/sigmod/GoldsteinL01}
\bibfield{author}{\bibinfo{person}{Jonathan Goldstein} {and}
  \bibinfo{person}{Per{-}{\AA}ke Larson}.} \bibinfo{year}{2001}\natexlab{}.
\newblock \showarticletitle{Optimizing Queries Using Materialized Views: {A}
  practical, scalable solution}. In \bibinfo{booktitle}{\emph{{SIGMOD}}}.
\newblock


\bibitem[\protect\citeauthoryear{Griffin and Libkin}{Griffin and
  Libkin}{1995}]%
        {DBLP:conf/sigmod/GriffinL95}
\bibfield{author}{\bibinfo{person}{Timothy Griffin} {and}
  \bibinfo{person}{Leonid Libkin}.} \bibinfo{year}{1995}\natexlab{}.
\newblock \showarticletitle{Incremental Maintenance of Views with Duplicates}.
  In \bibinfo{booktitle}{\emph{{SIGMOD}}}.
\newblock


\bibitem[\protect\citeauthoryear{Gupta, Agarwal, Tan, Kulesza, Pathak, Stefani,
  and Srinivasan}{Gupta et~al\mbox{.}}{2015}]%
        {DBLP:conf/sigmod/GuptaATKPSS15}
\bibfield{author}{\bibinfo{person}{Anurag Gupta}, \bibinfo{person}{Deepak
  Agarwal}, \bibinfo{person}{Derek Tan}, \bibinfo{person}{Jakub Kulesza},
  \bibinfo{person}{Rahul Pathak}, \bibinfo{person}{Stefano Stefani}, {and}
  \bibinfo{person}{Vidhya Srinivasan}.} \bibinfo{year}{2015}\natexlab{}.
\newblock \showarticletitle{Amazon Redshift and the Case for Simpler Data
  Warehouses}. In \bibinfo{booktitle}{\emph{{SIGMOD}}}.
\newblock


\bibitem[\protect\citeauthoryear{Gupta and Mumick}{Gupta and Mumick}{1995}]%
        {DBLP:journals/debu/GuptaM95}
\bibfield{author}{\bibinfo{person}{Ashish Gupta} {and}
  \bibinfo{person}{Inderpal~Singh Mumick}.} \bibinfo{year}{1995}\natexlab{}.
\newblock \showarticletitle{Maintenance of Materialized Views: Problems,
  Techniques, and Applications}.
\newblock \bibinfo{journal}{\emph{{IEEE} Data Eng. Bull.}}
  \bibinfo{volume}{18}, \bibinfo{number}{2} (\bibinfo{year}{1995}),
  \bibinfo{pages}{3--18}.
\newblock


\bibitem[\protect\citeauthoryear{Gupta}{Gupta}{1997}]%
        {DBLP:conf/icdt/Gupta97}
\bibfield{author}{\bibinfo{person}{Himanshu Gupta}.}
  \bibinfo{year}{1997}\natexlab{}.
\newblock \showarticletitle{Selection of Views to Materialize in a Data
  Warehouse}. In \bibinfo{booktitle}{\emph{{ICDT}}}.
\newblock


\bibitem[\protect\citeauthoryear{Gupta, Harinarayan, Rajaraman, and
  Ullman}{Gupta et~al\mbox{.}}{1997}]%
        {DBLP:conf/icde/GuptaHRU97}
\bibfield{author}{\bibinfo{person}{Himanshu Gupta}, \bibinfo{person}{Venky
  Harinarayan}, \bibinfo{person}{Anand Rajaraman}, {and}
  \bibinfo{person}{Jeffrey~D. Ullman}.} \bibinfo{year}{1997}\natexlab{}.
\newblock \showarticletitle{Index Selection for {OLAP}}. In
  \bibinfo{booktitle}{\emph{{ICDE}}}.
\newblock


\bibitem[\protect\citeauthoryear{Harinarayan, Rajaraman, and
  Ullman}{Harinarayan et~al\mbox{.}}{1996}]%
        {DBLP:conf/sigmod/HarinarayanRU96}
\bibfield{author}{\bibinfo{person}{Venky Harinarayan}, \bibinfo{person}{Anand
  Rajaraman}, {and} \bibinfo{person}{Jeffrey~D. Ullman}.}
  \bibinfo{year}{1996}\natexlab{}.
\newblock \showarticletitle{Implementing Data Cubes Efficiently}. In
  \bibinfo{booktitle}{\emph{{SIGMOD}}}.
\newblock


\bibitem[\protect\citeauthoryear{Heule, Nunkesser, and Hall}{Heule
  et~al\mbox{.}}{2013}]%
        {DBLP:conf/edbt/HeuleNH13}
\bibfield{author}{\bibinfo{person}{Stefan Heule}, \bibinfo{person}{Marc
  Nunkesser}, {and} \bibinfo{person}{Alexander Hall}.}
  \bibinfo{year}{2013}\natexlab{}.
\newblock \showarticletitle{HyperLogLog in practice: algorithmic engineering of
  a state of the art cardinality estimation algorithm}. In
  \bibinfo{booktitle}{\emph{{EDBT}}}.
\newblock


\bibitem[\protect\citeauthoryear{Huai, Chauhan, Gates, Hagleitner, Hanson,
  O'Malley, Pandey, Yuan, Lee, and Zhang}{Huai et~al\mbox{.}}{2014}]%
        {DBLP:conf/sigmod/HuaiCGHHOPYL014}
\bibfield{author}{\bibinfo{person}{Yin Huai}, \bibinfo{person}{Ashutosh
  Chauhan}, \bibinfo{person}{Alan Gates}, \bibinfo{person}{G{\"{u}}nther
  Hagleitner}, \bibinfo{person}{Eric~N. Hanson}, \bibinfo{person}{Owen
  O'Malley}, \bibinfo{person}{Jitendra Pandey}, \bibinfo{person}{Yuan Yuan},
  \bibinfo{person}{Rubao Lee}, {and} \bibinfo{person}{Xiaodong Zhang}.}
  \bibinfo{year}{2014}\natexlab{}.
\newblock \showarticletitle{Major technical advancements in apache hive}. In
  \bibinfo{booktitle}{\emph{{SIGMOD}}}.
\newblock


\bibitem[\protect\citeauthoryear{Isard, Budiu, Yu, Birrell, and Fetterly}{Isard
  et~al\mbox{.}}{2007}]%
        {DBLP:conf/eurosys/IsardBYBF07}
\bibfield{author}{\bibinfo{person}{Michael Isard}, \bibinfo{person}{Mihai
  Budiu}, \bibinfo{person}{Yuan Yu}, \bibinfo{person}{Andrew Birrell}, {and}
  \bibinfo{person}{Dennis Fetterly}.} \bibinfo{year}{2007}\natexlab{}.
\newblock \showarticletitle{Dryad: distributed data-parallel programs from
  sequential building blocks}. In \bibinfo{booktitle}{\emph{{EuroSys}}}.
\newblock


\bibitem[\protect\citeauthoryear{Jindal, Qiao, Patel, Yin, Di, Bag, Friedman,
  Lin, Karanasos, and Rao}{Jindal et~al\mbox{.}}{2018}]%
        {DBLP:conf/sigmod/JindalQPYDBFLKR18}
\bibfield{author}{\bibinfo{person}{Alekh Jindal}, \bibinfo{person}{Shi Qiao},
  \bibinfo{person}{Hiren Patel}, \bibinfo{person}{Zhicheng Yin},
  \bibinfo{person}{Jieming Di}, \bibinfo{person}{Malay Bag},
  \bibinfo{person}{Marc Friedman}, \bibinfo{person}{Yifung Lin},
  \bibinfo{person}{Konstantinos Karanasos}, {and} \bibinfo{person}{Sriram
  Rao}.} \bibinfo{year}{2018}\natexlab{}.
\newblock \showarticletitle{Computation Reuse in Analytics Job Service at
  Microsoft}. In \bibinfo{booktitle}{\emph{{SIGMOD}}}.
\newblock


\bibitem[\protect\citeauthoryear{Kornacker, Behm, Bittorf, Bobrovytsky, Ching,
  Choi, Erickson, Grund, Hecht, Jacobs, Joshi, Kuff, Kumar, Leblang, Li,
  Pandis, Robinson, Rorke, Rus, Russell, Tsirogiannis, Wanderman{-}Milne, and
  Yoder}{Kornacker et~al\mbox{.}}{2015}]%
        {DBLP:conf/cidr/KornackerBBBCCE15}
\bibfield{author}{\bibinfo{person}{Marcel Kornacker},
  \bibinfo{person}{Alexander Behm}, \bibinfo{person}{Victor Bittorf},
  \bibinfo{person}{Taras Bobrovytsky}, \bibinfo{person}{Casey Ching},
  \bibinfo{person}{Alan Choi}, \bibinfo{person}{Justin Erickson},
  \bibinfo{person}{Martin Grund}, \bibinfo{person}{Daniel Hecht},
  \bibinfo{person}{Matthew Jacobs}, \bibinfo{person}{Ishaan Joshi},
  \bibinfo{person}{Lenni Kuff}, \bibinfo{person}{Dileep Kumar},
  \bibinfo{person}{Alex Leblang}, \bibinfo{person}{Nong Li},
  \bibinfo{person}{Ippokratis Pandis}, \bibinfo{person}{Henry Robinson},
  \bibinfo{person}{David Rorke}, \bibinfo{person}{Silvius Rus},
  \bibinfo{person}{John Russell}, \bibinfo{person}{Dimitris Tsirogiannis},
  \bibinfo{person}{Skye Wanderman{-}Milne}, {and} \bibinfo{person}{Michael
  Yoder}.} \bibinfo{year}{2015}\natexlab{}.
\newblock \showarticletitle{Impala: {A} Modern, Open-Source {SQL} Engine for
  Hadoop}. In \bibinfo{booktitle}{\emph{{CIDR}}}.
\newblock


\bibitem[\protect\citeauthoryear{Kreps, Narkhede, and Rao}{Kreps
  et~al\mbox{.}}{2011}]%
        {DBLP:Kreps2011Kafka}
\bibfield{author}{\bibinfo{person}{Jay Kreps}, \bibinfo{person}{Neha Narkhede},
  {and} \bibinfo{person}{Jun Rao}.} \bibinfo{year}{2011}\natexlab{}.
\newblock \showarticletitle{Kafka : a Distributed Messaging System for Log
  Processing}. In \bibinfo{booktitle}{\emph{{NetDB}}}.
\newblock


\bibitem[\protect\citeauthoryear{Kubernetes}{Kubernetes}{2018}]%
        {website:Kubernetes}
Kubernetes \bibinfo{year}{2018}\natexlab{}.
\newblock \bibinfo{title}{{Kubernetes}: Production-Grade Container
  Orchestration}.
\newblock
\newblock
\newblock
\shownote{http://kubernetes.io/.}


\bibitem[\protect\citeauthoryear{Larson, Clinciu, Fraser, Hanson, Mokhtar,
  Nowakiewicz, Papadimos, Price, Rangarajan, Rusanu, and Saubhasik}{Larson
  et~al\mbox{.}}{2013}]%
        {DBLP:conf/sigmod/LarsonCFHMNPPRRS13}
\bibfield{author}{\bibinfo{person}{Per{-}{\AA}ke Larson},
  \bibinfo{person}{Cipri Clinciu}, \bibinfo{person}{Campbell Fraser},
  \bibinfo{person}{Eric~N. Hanson}, \bibinfo{person}{Mostafa Mokhtar},
  \bibinfo{person}{Michal Nowakiewicz}, \bibinfo{person}{Vassilis Papadimos},
  \bibinfo{person}{Susan~L. Price}, \bibinfo{person}{Srikumar Rangarajan},
  \bibinfo{person}{Remus Rusanu}, {and} \bibinfo{person}{Mayukh Saubhasik}.}
  \bibinfo{year}{2013}\natexlab{}.
\newblock \showarticletitle{Enhancements to {SQL} server column stores}. In
  \bibinfo{booktitle}{\emph{{SIGMOD}}}.
\newblock


\bibitem[\protect\citeauthoryear{Melnik, Gubarev, Long, Romer, Shivakumar,
  Tolton, and Vassilakis}{Melnik et~al\mbox{.}}{2010}]%
        {DBLP:journals/pvldb/MelnikGLRSTV10}
\bibfield{author}{\bibinfo{person}{Sergey Melnik}, \bibinfo{person}{Andrey
  Gubarev}, \bibinfo{person}{Jing~Jing Long}, \bibinfo{person}{Geoffrey Romer},
  \bibinfo{person}{Shiva Shivakumar}, \bibinfo{person}{Matt Tolton}, {and}
  \bibinfo{person}{Theo Vassilakis}.} \bibinfo{year}{2010}\natexlab{}.
\newblock \showarticletitle{Dremel: Interactive Analysis of Web-Scale
  Datasets}. In \bibinfo{booktitle}{\emph{{PVLDB}}}.
\newblock


\bibitem[\protect\citeauthoryear{O'Neil, O'Neil, Chen, and Revilak}{O'Neil
  et~al\mbox{.}}{2009}]%
        {DBLP:conf/tpctc/ONeilOCR09}
\bibfield{author}{\bibinfo{person}{Patrick~E. O'Neil},
  \bibinfo{person}{Elizabeth~J. O'Neil}, \bibinfo{person}{Xuedong Chen}, {and}
  \bibinfo{person}{Stephen Revilak}.} \bibinfo{year}{2009}\natexlab{}.
\newblock \showarticletitle{The Star Schema Benchmark and Augmented Fact Table
  Indexing}. In \bibinfo{booktitle}{\emph{{TPCTC}}}.
\newblock


\bibitem[\protect\citeauthoryear{Presto}{Presto}{2018}]%
        {website:Presto}
Presto \bibinfo{year}{2018}\natexlab{}.
\newblock \bibinfo{title}{{Presto}: Distributed SQL query engine for big data}.
\newblock
\newblock
\newblock
\shownote{http://prestodb.io/.}


\bibitem[\protect\citeauthoryear{Redshift}{Redshift}{2018}]%
        {website:Redshift}
Redshift \bibinfo{year}{2018}\natexlab{}.
\newblock \bibinfo{title}{{Amazon Redshift}: Amazon Web Services}.
\newblock
\newblock
\newblock
\shownote{http://aws.amazon.com/redshift/.}


\bibitem[\protect\citeauthoryear{Saha, Shah, Seth, Vijayaraghavan, Murthy, and
  Curino}{Saha et~al\mbox{.}}{2015}]%
        {DBLP:conf/sigmod/SahaSSVMC15}
\bibfield{author}{\bibinfo{person}{Bikas Saha}, \bibinfo{person}{Hitesh Shah},
  \bibinfo{person}{Siddharth Seth}, \bibinfo{person}{Gopal Vijayaraghavan},
  \bibinfo{person}{Arun~C. Murthy}, {and} \bibinfo{person}{Carlo Curino}.}
  \bibinfo{year}{2015}\natexlab{}.
\newblock \showarticletitle{Apache Tez: {A} Unifying Framework for Modeling and
  Building Data Processing Applications}. In
  \bibinfo{booktitle}{\emph{{SIGMOD}}}.
\newblock


\bibitem[\protect\citeauthoryear{Snowflake}{Snowflake}{2018}]%
        {website:Snowflake}
Snowflake \bibinfo{year}{2018}\natexlab{}.
\newblock \bibinfo{title}{{Snowflake}: The Enterprise Data Warehouse Built in
  the Cloud}.
\newblock
\newblock
\newblock
\shownote{http://www.snowflake.com/.}


\bibitem[\protect\citeauthoryear{Soliman, Antova, Raghavan, El{-}Helw, Gu,
  Shen, Caragea, Garcia{-}Alvarado, Rahman, Petropoulos, Waas, Narayanan,
  Krikellas, and Baldwin}{Soliman et~al\mbox{.}}{2014}]%
        {DBLP:conf/sigmod/SolimanAREGSCGRPWNKB14}
\bibfield{author}{\bibinfo{person}{Mohamed~A. Soliman},
  \bibinfo{person}{Lyublena Antova}, \bibinfo{person}{Venkatesh Raghavan},
  \bibinfo{person}{Amr El{-}Helw}, \bibinfo{person}{Zhongxian Gu},
  \bibinfo{person}{Entong Shen}, \bibinfo{person}{George~C. Caragea},
  \bibinfo{person}{Carlos Garcia{-}Alvarado}, \bibinfo{person}{Foyzur Rahman},
  \bibinfo{person}{Michalis Petropoulos}, \bibinfo{person}{Florian Waas},
  \bibinfo{person}{Sivaramakrishnan Narayanan}, \bibinfo{person}{Konstantinos
  Krikellas}, {and} \bibinfo{person}{Rhonda Baldwin}.}
  \bibinfo{year}{2014}\natexlab{}.
\newblock \showarticletitle{Orca: a modular query optimizer architecture for
  big data}. In \bibinfo{booktitle}{\emph{{SIGMOD}}}.
\newblock


\bibitem[\protect\citeauthoryear{Stillger, Lohman, Markl, and Kandil}{Stillger
  et~al\mbox{.}}{2001}]%
        {DBLP:conf/vldb/StillgerLMK01}
\bibfield{author}{\bibinfo{person}{Michael Stillger}, \bibinfo{person}{Guy~M.
  Lohman}, \bibinfo{person}{Volker Markl}, {and} \bibinfo{person}{Mokhtar
  Kandil}.} \bibinfo{year}{2001}\natexlab{}.
\newblock \showarticletitle{{LEO} - DB2's LEarning Optimizer}. In
  \bibinfo{booktitle}{\emph{{PVLDB}}}.
\newblock


\bibitem[\protect\citeauthoryear{Stonebraker and {\c{C}}etintemel}{Stonebraker
  and {\c{C}}etintemel}{2005}]%
        {DBLP:conf/icde/StonebrakerC05}
\bibfield{author}{\bibinfo{person}{Michael Stonebraker} {and}
  \bibinfo{person}{Ugur {\c{C}}etintemel}.} \bibinfo{year}{2005}\natexlab{}.
\newblock \showarticletitle{"One Size Fits All": An Idea Whose Time Has Come
  and Gone}. In \bibinfo{booktitle}{\emph{{ICDE}}}.
\newblock


\bibitem[\protect\citeauthoryear{Thusoo, Sarma, Jain, Shao, Chakka, Zhang,
  Anthony, Liu, and Murthy}{Thusoo et~al\mbox{.}}{2010}]%
        {DBLP:conf/icde/ThusooSJSCZALM10}
\bibfield{author}{\bibinfo{person}{Ashish Thusoo}, \bibinfo{person}{Joydeep~Sen
  Sarma}, \bibinfo{person}{Namit Jain}, \bibinfo{person}{Zheng Shao},
  \bibinfo{person}{Prasad Chakka}, \bibinfo{person}{Ning Zhang},
  \bibinfo{person}{Suresh Anthony}, \bibinfo{person}{Hao Liu}, {and}
  \bibinfo{person}{Raghotham Murthy}.} \bibinfo{year}{2010}\natexlab{}.
\newblock \showarticletitle{Hive - a petabyte scale data warehouse using
  Hadoop}. In \bibinfo{booktitle}{\emph{{ICDE}}}.
\newblock


\bibitem[\protect\citeauthoryear{Vavilapalli, Murthy, Douglas, Agarwal, Konar,
  Evans, Graves, Lowe, Shah, Seth, Saha, Curino, O'Malley, Radia, Reed, and
  Baldeschwieler}{Vavilapalli et~al\mbox{.}}{2013}]%
        {DBLP:conf/cloud/VavilapalliMDAKEGLSSSCORRB13}
\bibfield{author}{\bibinfo{person}{Vinod~Kumar Vavilapalli},
  \bibinfo{person}{Arun~C. Murthy}, \bibinfo{person}{Chris Douglas},
  \bibinfo{person}{Sharad Agarwal}, \bibinfo{person}{Mahadev Konar},
  \bibinfo{person}{Robert Evans}, \bibinfo{person}{Thomas Graves},
  \bibinfo{person}{Jason Lowe}, \bibinfo{person}{Hitesh Shah},
  \bibinfo{person}{Siddharth Seth}, \bibinfo{person}{Bikas Saha},
  \bibinfo{person}{Carlo Curino}, \bibinfo{person}{Owen O'Malley},
  \bibinfo{person}{Sanjay Radia}, \bibinfo{person}{Benjamin Reed}, {and}
  \bibinfo{person}{Eric Baldeschwieler}.} \bibinfo{year}{2013}\natexlab{}.
\newblock \showarticletitle{Apache Hadoop {YARN:} yet another resource
  negotiator}. In \bibinfo{booktitle}{\emph{{SOCC}}}.
\newblock


\bibitem[\protect\citeauthoryear{Wiederhold}{Wiederhold}{1992}]%
        {DBLP:journals/computer/Wiederhold92}
\bibfield{author}{\bibinfo{person}{Gio Wiederhold}.}
  \bibinfo{year}{1992}\natexlab{}.
\newblock \showarticletitle{Mediators in the Architecture of Future Information
  Systems}.
\newblock \bibinfo{journal}{\emph{{IEEE} Computer}} \bibinfo{volume}{25},
  \bibinfo{number}{3} (\bibinfo{year}{1992}), \bibinfo{pages}{38--49}.
\newblock


\bibitem[\protect\citeauthoryear{Yang, Tschetter, L{\'{e}}aut{\'{e}}, Ray,
  Merlino, and Ganguli}{Yang et~al\mbox{.}}{2014}]%
        {DBLP:conf/sigmod/YangTLRMG14}
\bibfield{author}{\bibinfo{person}{Fangjin Yang}, \bibinfo{person}{Eric
  Tschetter}, \bibinfo{person}{Xavier L{\'{e}}aut{\'{e}}},
  \bibinfo{person}{Nelson Ray}, \bibinfo{person}{Gian Merlino}, {and}
  \bibinfo{person}{Deep Ganguli}.} \bibinfo{year}{2014}\natexlab{}.
\newblock \showarticletitle{Druid: a real-time analytical data store}. In
  \bibinfo{booktitle}{\emph{{SIGMOD}}}.
\newblock


\bibitem[\protect\citeauthoryear{Zaharia, Chowdhury, Franklin, Shenker, and
  Stoica}{Zaharia et~al\mbox{.}}{2010}]%
        {DBLP:conf/hotcloud/ZahariaCFSS10}
\bibfield{author}{\bibinfo{person}{Matei Zaharia}, \bibinfo{person}{Mosharaf
  Chowdhury}, \bibinfo{person}{Michael~J. Franklin}, \bibinfo{person}{Scott
  Shenker}, {and} \bibinfo{person}{Ion Stoica}.}
  \bibinfo{year}{2010}\natexlab{}.
\newblock \showarticletitle{Spark: Cluster Computing with Working Sets}. In
  \bibinfo{booktitle}{\emph{{USENIX} HotCloud}}.
\newblock


\bibitem[\protect\citeauthoryear{Za{\"{\i}}t, Chakkappen, Budalakoti, Valluri,
  Krishnamachari, and Wood}{Za{\"{\i}}t et~al\mbox{.}}{2017}]%
        {DBLP:journals/pvldb/ZaitCBVKW17}
\bibfield{author}{\bibinfo{person}{Mohamed Za{\"{\i}}t}, \bibinfo{person}{Sunil
  Chakkappen}, \bibinfo{person}{Suratna Budalakoti},
  \bibinfo{person}{Satyanarayana~R. Valluri}, \bibinfo{person}{Ramarajan
  Krishnamachari}, {and} \bibinfo{person}{Alan Wood}.}
  \bibinfo{year}{2017}\natexlab{}.
\newblock \showarticletitle{Adaptive Statistics in Oracle 12c}. In
  \bibinfo{booktitle}{\emph{{PVLDB}}}.
\newblock


\bibitem[\protect\citeauthoryear{Zilio, Rao, Lightstone, Lohman, Storm,
  Garcia{-}Arellano, and Fadden}{Zilio et~al\mbox{.}}{2004}]%
        {DBLP:conf/vldb/ZilioRLLSGF04}
\bibfield{author}{\bibinfo{person}{Daniel~C. Zilio}, \bibinfo{person}{Jun Rao},
  \bibinfo{person}{Sam Lightstone}, \bibinfo{person}{Guy~M. Lohman},
  \bibinfo{person}{Adam~J. Storm}, \bibinfo{person}{Christian
  Garcia{-}Arellano}, {and} \bibinfo{person}{Scott Fadden}.}
  \bibinfo{year}{2004}\natexlab{}.
\newblock \showarticletitle{{DB2} Design Advisor: Integrated Automatic Physical
  Database Design}. In \bibinfo{booktitle}{\emph{{PVLDB}}}.
\newblock


\end{thebibliography}

\end{document}